\documentclass[aps,
 prx,amsmath,amssymb,showpacs,showkeys,longbibliography,
 reprint,
]{revtex4-1}

\usepackage{graphicx}
\usepackage{dcolumn}
\usepackage{bm}
\usepackage{color}

\begin{document}


\title[Sample title]{Practical quantum realization of the ampere from the elementary charge}

\author{J. Brun-Picard, S. Djordjevic, D. Leprat, F. Schopfer, W. Poirier}
\email{wilfrid.poirier@lne.fr}
\affiliation{LNE - Laboratoire national de m\'{e}trologie et d'essais, 78197 Trappes, France}

\date{\today}

\begin{abstract}
One major change of the future revision of the International System of Units (SI) is a new definition of the ampere based on the elementary charge \emph{e}. Replacing the former definition based on Amp\`ere's force law will allow one to fully benefit from quantum physics to realize the ampere. However, a quantum realization of the ampere from \emph{e}, accurate to within $10^{-8}$ in relative value and fulfilling traceability needs, is still missing despite many efforts have been spent for the development of single-electron tunneling devices. Starting again with Ohm's law, applied here in a quantum circuit combining the quantum Hall resistance and Josephson voltage standards with a superconducting cryogenic amplifier, we report on a practical and universal programmable quantum current generator. We demonstrate that currents generated in the milliampere range are quantized in terms of $ef_\mathrm{J}$ ($f_\mathrm{J}$ is the Josephson frequency) with a measurement uncertainty of $10^{-8}$. This new quantum current source, able to deliver such accurate currents down to the microampere range, can greatly improve the current measurement traceability, as demonstrated with the calibrations of digital ammeters. Beyond, it opens the way to further developments in metrology and in fundamental physics, such as a quantum multimeter or new accurate comparisons to single electron pumps.
\end{abstract}

\maketitle
\section{\label{sec:level1}Introduction}
Measurements rely on the International System of Units (SI)\cite{BrochureSI} which is a consistent system constructed historically on seven based units namely the meter (m), the kilogram (kg), the second (s), the ampere (A), the kelvin (K), the mole (mol) and the candela (cd), all other units being formed as products of powers of the base units. The SI has always evolved following the scientific knowledge with the aim to decrease the uncertainty in the measurements, but also with the aim of universality. This is best illustrated by the history of the definition of the meter which was first based on an artefact then on a reference to an atomic transition and more recently related to the second through a fixed value of the speed of light $c$ expressed in the unit m.s$^{-1}$. This success has guided the choice for the future revision of the SI \cite{CGPM2014,Mills2005,Mills2006,Milton2007,Draft2015}, in which the definitions of the seven base units will be based on constants ranging from fundamental constants of nature to technical constants \cite{Draft2015}. Quantum mechanics will be fully exploited by fixing the values of the Planck constant \emph{h} and of the elementary charge \emph{e}. From the definitions of the second and the meter, these fundamental constants  expressed in the units $~\mathrm{kg.m^2.s^{-1}}$ and $\mathrm{A.s}$ respectively will set the definitions of the kilogram\cite{Mills2005} and of the ampere\cite{Mills2006} without specifying the experiment for their realizations. The mass unit, bound to \emph{h}, will be no more realized by the International prototype of the kilogram suspected drifting with time but for example using the watt balance experiment\cite{Kibble1976,Stock2013}. Similarly, the ampere will be realized from the elementary charge \emph{e} and the frequency $f (\mathrm{s}^{-1})$, and no longer from Amp\`ere's force law\cite{AmpereDefinition} which relates electrical units to mechanical units and thereby limits the relative uncertainty to a few parts in $10^{7}$\cite{SIANNEXE2}.

 A direct way to realize the ampere from the elementary charge and the frequency $f$, illustrated in Fig.1a, is based on single-electron tunneling (SET) devices\cite{Pekola2013} in mesoscopic systems at very low temperatures where the charge quantization manifests itself due to Coulomb blockade \cite{Grabert1991}. Among SET devices, electron pumps\cite{Pothier1992} transfer a precise number $n_{Q}$ of charge $Q\equiv e$ at each cycle of a control parameter which is synchronized to an external frequency $f_\mathrm{P}$, so that the amplitude of the output current is ideally equal to $n_{Q}Qf_\mathrm{P}$, \emph{i.e} theoretically equal to $n_{Q}ef_\mathrm{P}$. The first electron pumps consisted of small metallic islands in series isolated by tunnel junctions. In a 7-junction device, an error rate per cycle of $1.5\times10^{-8}$ was measured for frequencies in the MHz range\cite{Keller1996}. The accuracy of such devices, obtained by charging a cryogenic capacitor with a precise number of electrons\cite{Keller1999}, was then demonstrated with a relative uncertainty of $9.2\times10^{-7}$ for currents below 1 pA\cite{Keller2007}. A similar experiment reached a relative uncertainty of $1.66\times10^{-6}$ with a 5-junction R-pump\cite{Camarota2012} (Fig.1b). Recently, alternative electron pumps based on tunable barriers in a non-adiabatic regime were proposed as a trade-off between accuracy and increased current, as reviewed in references \cite{Pekola2013,Kaestner2015}. In most recent devices operating at very low temperatures ($T\leq 0.3~\mathrm{K}$) and under high magnetic fields ($B\geq 14~\mathrm{T}$), the quantization of the current was demonstrated with relative measurement uncertainties of $1.2\times10^{-6}$ at 150 pA ($f_\mathrm{P}=945~\mathrm{MHz}$)\cite{Giblin2012} and $2\times10^{-7}$ at 90 pA ($f_\mathrm{P}=545~\mathrm{MHz}$)\cite{Stein2015} (Fig.1b).
\begin{figure*}[t]
\includegraphics[width=17cm]{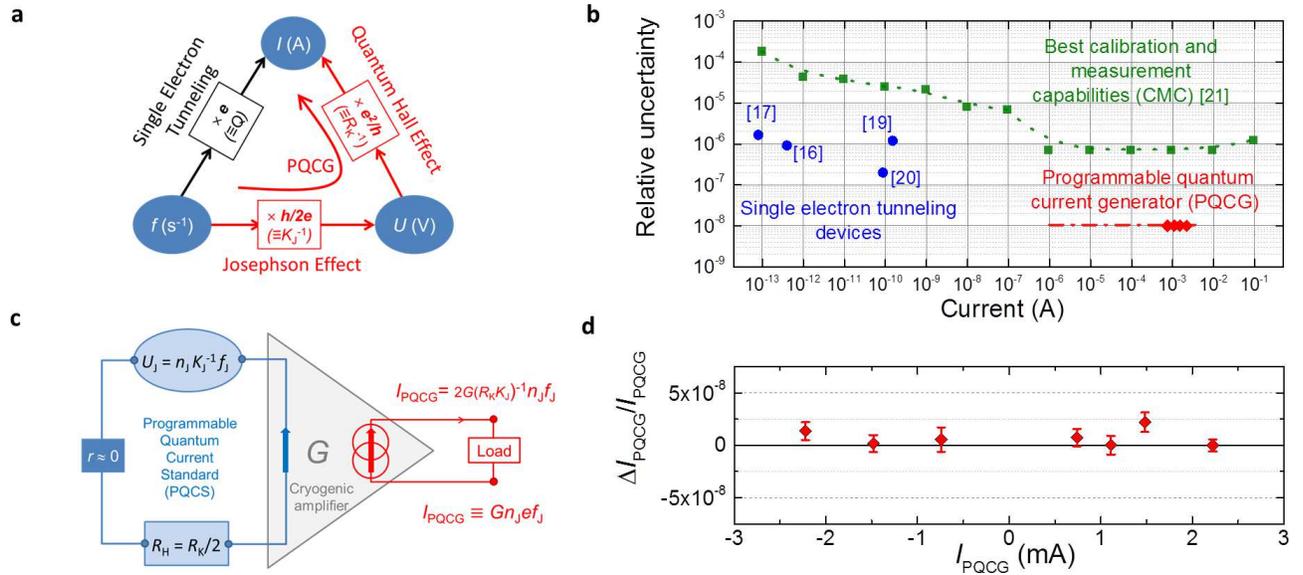}\hfill
\caption{\textbf{Practical realization of the ampere from quantum electrical effects.} a) Illustration of the two ways to generate a current from the frequency $f (\mathrm{s}^{-1})$, expressed in ampere in the future SI in terms of the elementary charge $e$. The current can be generated either using single electron tunneling devices or the combination of the Josephson effect and the quantum Hall effect. Quantum effects involve fundamental constants whose uncertainties in the present SI will be reduced to zero in the future SI. b) Comparison of the relative uncertainty of the current generated by the PQCG in the milliampere range (red diamonds) to state-of-the-art current measurements/generation. These high accuracy results include the best calibration and measurement capabilities (CMCs) (green squares: from $10^{-13}$ up to $10^{-11}$ A by charging a capacitor (Physikalisch-Technische Bundesanstalt), from $10^{-10}$ up to $10^{-1}$ A by applying Ohm's law (Laboratoire national de m\'{e}trologie et d'essais))\cite{BestCMC} and the main SET devices results mentioned in the text (blue dots). The red dashed-dotted line shows the estimated relative uncertainties of the current generated by the PQCG from $\mathrm{1~\mu A}$ up to 10 mA. c) Principle of the programmable quantum current generator (PQCG). d) Relative deviation $\Delta I_\mathrm{PQCG}/I_\mathrm{PQCG}$ of the measured current from the quantized value in the milliampere range. Error bars are combined standard uncertainties (1 s.d.). No significant relative discrepancies can be observed within an uncertainty of $10^{-8}$.}\label{fig:Fig1}
\end{figure*}

As illustrated in Fig.1a, the future definition of the ampere can also be realized by applying Ohm's law to the quantum voltage and resistance standards that are based on the Josephson effect (JE)\cite{Josephson62}, and the quantum Hall effect (QHE)\cite{Klitzing1980}, two gauge-invariant macroscopic quantum effects that involve the Josephson and the von Klitzing constants, $K{_\mathrm J} \equiv 2e/h$ and $R{_\mathrm K} \equiv h/e^2$ respectively. More precisely, the AC (alternative current) JE converts the frequency $f$ of an electromagnetic wave to a voltage $U$ with the constant $K{_\mathrm J}^{-1}$. This effect is characterized by the appearance of quantized voltage steps (Shapiro steps)\cite{Shapiro63}, at values $n_\mathrm{J}K{_\mathrm J}^{-1}f_\mathrm{J}$ in the current-voltage characteristic of a Josephson junction irradiated by a microwave field of frequency $f_\mathrm{J}$, where $n_\mathrm{J}$ is an integer. This can be understood as the transfer of an integer number of flux quanta $\phi_0=h/2e$ per microwave cycle due to the circulation of a current of Cooper pairs. The QHE links the current $I$ to the voltage $U$ through the constant $R{_\mathrm K}^{-1}$. This quantum phenomenon manifests itself in a device based on a two-dimensional electron gas under a perpendicular magnetic field, by the quantization of the Hall resistance at values $R_\mathrm{K}/i_\mathrm{K}$, where $i_\mathrm{K}$ is an integer. This comes from the existence at Fermi energy of $i_\mathrm{K}$ one-dimensional ballistic chiral states\cite{Buttiker1988} of conductance $e^2/h$ each at the device edges, and the absence of delocalized states in the bulk due to the opening of a gap in the energy spectrum (the density of states is quantized in Landau levels). From the application of Ohm's law to these two quantum standards, the frequency $f$ can therefore be converted in a current $I$ with the constant $(K_{\mathrm J}\times R_{\mathrm K})^{-1}\equiv e/2$.
It is planned that the relationships $K_\mathrm{J}=2e/h$ and $R_\mathrm{K}=h/e^2$ will be assumed in the new SI. This is a reasonable assumption since no measurable deviation has been predicted by quantum mechanics\cite{Bloch1970,Laughlin81,Thouless1994,Penin2009,Penin2010} and no significant deviation has been demonstrated experimentally, either by independent determinations of the constants in the relations\cite{MohrCODATA2010,codata16,Pekola2013} or by universality tests. Several experiments have indeed demonstrated the universality of the JE and the QHE with relative measurements uncertainties below $2\times10^{-16}$\cite{Tsai1983,Jain1987,Krasnopolin2002} and $10^{-10}$\cite{Schopfer2013,Ribeiro2015,Janssen2011} respectively. In the new SI, the Josephson voltage standard (JVS) and the quantum Hall resistance standard (QHRS) would therefore become realizations of the ohm and the volt with relative uncertainties below $10^{-9}$, only limited by their experimental implementation, and no longer by the uncertainties on $K_\mathrm{J}$ and $R_\mathrm{K}$ of $4\times10^{-7}$\cite{KJ} and $1\times10^{-7}$\cite{RK} respectively, in the present SI (Appendix A). Applying Ohm's law to those standards in the new SI would result in a current standard which can be expressed as $i_\mathrm{K}n_\mathrm{J}ef_\mathrm{J}/2$, where $e$ has an exact value. The expectation is to reach an accurate quantum current standard realizing the ampere definition with a high level of reproducibility and universality, which directly benefits from that of the JVS and the QHRS. This goal was unattainable with the former definition established in 1948. Nowadays, National Metrology Institutes already apply Ohm's law to secondary voltage and resistance standards, traceable to $K_\mathrm{J}$ and $R_\mathrm{K}$, for the current traceability. However, the uncertainty claimed in their calibration and measurement capabilities (CMC) for current, reported in Fig.1b, is in practice not better than $10^{-6}$\cite{BestCMC}. Above $\mathrm{1~\mu A}$, this limitation is mainly caused by higher calibration uncertainties of secondary standards and the lack of accurate and stable true current sources, while below this current limit, the uncertainty which increases steadily towards lower current levels is rather due a lack of sensitivity of the measurement techniques.
These CMCs emphasize the advantage of an accurate reference current standard able to deliver high currents in order to optimize and shorten the current traceability in NMIs over a wide range of values above $\mathrm{1~\mu A}$. In the range of lower currents, a traceability improvement might be expected by exploiting not only SET devices as quantum current standards but also accurate current amplifiers to make the link between low currents and higher current references. A recent example of low-noise amplifiers that operates at room temperature, is an ultra low-noise current amplifier (ULCA) \cite{Stein2015,DrungRSI2015,Drung2015}, stable within $10^{-7}$ over one week and having a typical long term drift of $5\times10^{-6}$ per year. The downscaling approach also further motivates the development of current standards with large values.

Here, we report on a programmable quantum current generator (PQCG), linked to the elementary charge \emph{e}, which is built from an application of Ohm's law to quantum standards combined in an original quantum circuit (Fig.1c). Shortly, it is based on a current source locked, by means of an highly-accurate cryogenic amplifier of gain $G$ using a magnetic coupling, to a multiple or fraction value of a programmable quantum current standard (PQCS) used as a reference. The PQCS is the current circulating in a closed circuit formed by a Josephson voltage $U_\mathrm{J}=n_\mathrm{J}K_\mathrm{J}^{-1}f_\mathrm{J}$ applied to a quantum Hall resistance standard $R_\mathrm{H}=R_\mathrm{K}/2$ using a special connection scheme which drastically reduces the two-wires series resistance ($r\simeq 0$) thanks to the QHE properties and which allows its accurate detection by the amplifier. Fig. 1d demonstrates that currents generated by the PQCG from $\pm 0.7$ to $\pm 2.2 \mathrm{mA}$ are perfectly quantized in terms of $(K_\mathrm{J}\times R_\mathrm{K})^{-1}\equiv e/2$, within a combined standard uncertainty (Appendix B) of $10^{-8}$ (one standard deviation or 1 s.d.). By principle, this new standard is programmable and versatile, \emph{i.e.} it can generate currents over a wide range of values, extending from 10 mA down to $\mathrm{1~\mu A}$. The PQCG uncertainty budget reported in Fig.1b by the red dashed-dotted line shows that the accuracy is unchanged in the whole current range. The PQCG is a primary quantum current standard, accurate over a wide current range, able to greatly improve the current measurement traceability by reducing uncertainties of two orders of magnitude compared to those declared in best CMCs. This is demonstrated by the calibration of a digital ammeter (DA) on several current ranges from $\mathrm{1~\mu A}$ to 5 mA with measurement uncertainties only limited by the device under test. More fundamentally, the PQCG is able to implement the future ampere definition in terms of the elementary charge \emph{e} with the target uncertainty of $10^{-8}$. This will rely on the adoption of the solid-state quantum theory in the planned new SI. In this context, demonstrating the equivalence of the two quantum realizations of the ampere described in Fig.1a, so-called closing the metrological triangle\cite{Likharev1985}, is a challenging experiment of great interest. Improvements in its realization are expected from the PQCG and the quantum circuit methods reported here.
\begin{figure*}[t]
\includegraphics[width=17cm]{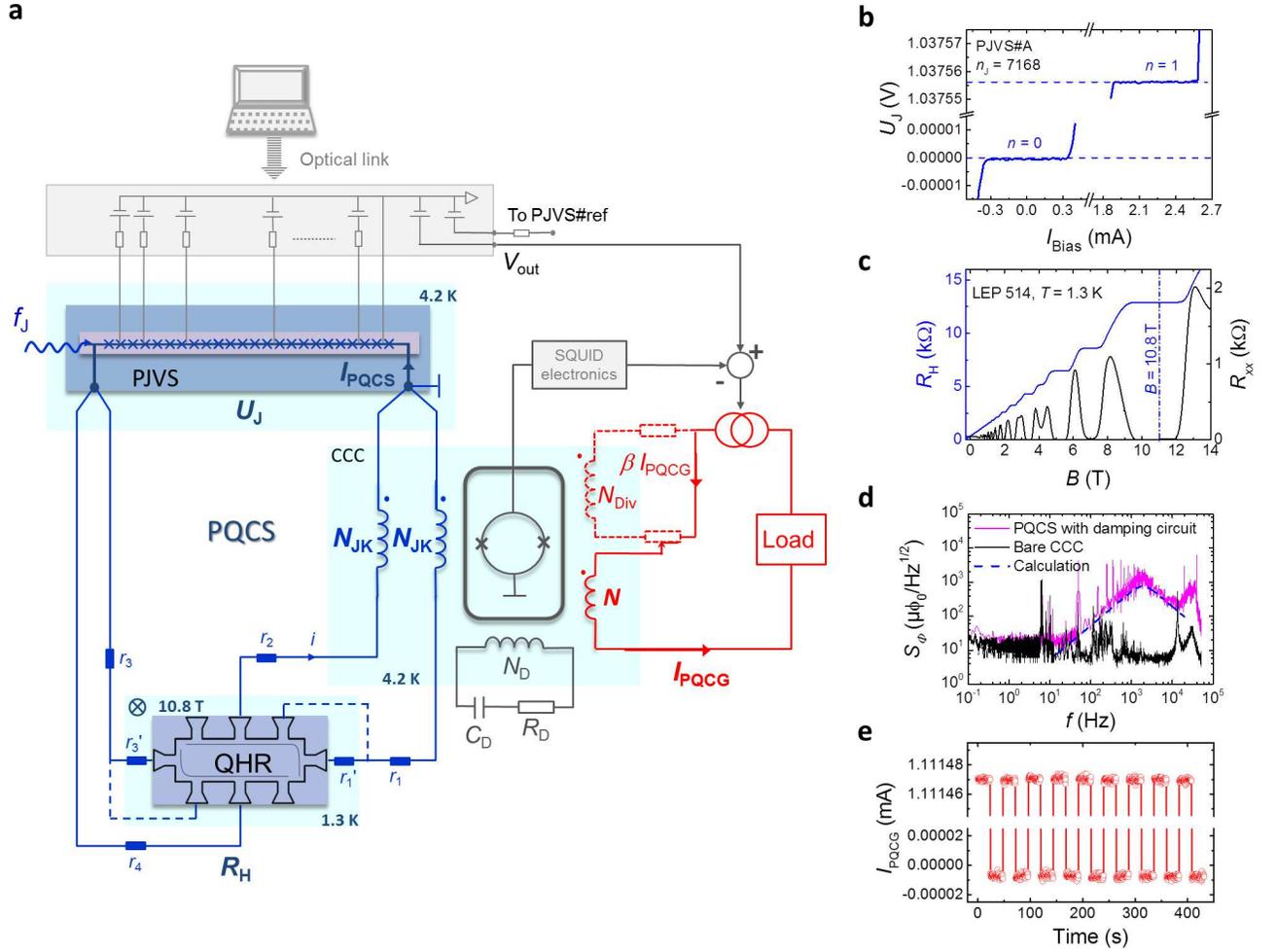}\hfill
\caption{\textbf{Experimental realization of the PQCG.} a) The PQCS is composed of a PJVS biasing a QHRS through double connections, each one incorporating a cryogenic current comparator (CCC) winding ($N_\mathrm{JK}$ turns) on the low potential side. A third terminal (dotted line) connected at the top of the QHE setup further reduces the cable contribution to the current $I_\mathrm{PQCS}$. The current $I_\mathrm{PQCG}$ of the PQCG, generated by an external current source into a winding of $N$ turns, is synchronized and coarsely adjusted by the PJVS programmable bias source using $V_\mathrm{out}$. $I_\mathrm{PQCG}$ is locked to the current $I_\mathrm{PQCS}$ of the PQCS by means of the CCC which is used as an accurate adder-amplifier of $N_\mathrm{JK}/N$ gain. A current divider injecting a fraction $\beta$ of $I_\mathrm{PQCG}$ in a CCC winding of $N_\mathrm{Div}$(=16~turns) allows a fine tuning of the CCC amplification gain. The damping circuit formed by a 100 nF highly-insulated Polytetrafluoroethylene (PTFE) capacitance in series with a $\mathrm{1.1~k\Omega}$ resistor is connected to a CCC winding of $N_\mathrm{D}$(=1600) turns. b) PJVS\#A output voltage as a function of the bias current $I_\mathrm{Bias}$ for the total number of Josephson junctions at $f_\mathrm{J}=70$ GHz and \emph{T}=4.2 K. c) Hall resistance $R_{\mathrm H}$ and longitudinal resistance $R_{xx}$ measured, using a current of $10~\mathrm{\mu A}$ at \emph{T}=1.3 K, as a function of \emph{B} in the GaAs/AlGaAs-based Hall bar device (LEP514). The device is used at \emph{B}=10.8 T. d) The noise amplitude spectral density (expressed in $\mathrm{\mu \phi_0/Hz^{1/2}}$), measured at the output of the SQUID in internal feedback mode, for the CCC alone (black) and the CCC connected to the PQCS and the damping circuit (magenta) without the external current source. e) Series of on-off switchings of the current $I_\mathrm{PQCG}$ recorded on a digital ammeter (at $1.1~\mathrm{mA}$), obtained for $n_\mathrm{J}=3073$, $N_\mathrm{JK}=129$ and $N=4$.}\label{fig:Fig2}
\end{figure*}

\section{\label{sec:level1}The programmable quantum current generator}
\subsection{\label{sec:level1}Realization}
The experimental scheme of the PQCG, described in Fig.2a, aims at realizing the principle shown in Fig.1c. The PQCS is built from a programmable Josephson voltage standard (PJVS)\cite{BehrMST2012} which is used to maintain the quantized voltage at the terminals of a quantum Hall resistance standard (QHRS)\cite{Jeckelmann2001,Poirier2009} of resistance $R_{\mathrm H} = R_\mathrm{K}/2 \equiv h/2e^2$. The PJVS is based on a 1 V series array of SINIS Josephson junctions\cite{Mueller2007}, where S, I and N correspond to superconductor, insulator and normal metal respectively, operating at frequencies $f_{\mathrm J}\sim70~\mathrm{GHz}$. The array is divided in segments that can be individually biased on the $n=0$ or $n=\pm 1$ Shapiro steps by a programmable bias source. The quantized voltage steps are given by $U_{\mathrm J} = \pm n_{\mathrm J} (K_\mathrm{J}^{-1}) f_{\mathrm J}\equiv \pm n_{\mathrm J} (h/2e) f_{\mathrm J}$, where $n_{\mathrm J}$ is now the number of biased junctions on the first Shapiro step and which can be as large as several thousands (Fig.2b and Appendix C.1). The current $I_\mathrm{PQCS}$ circulating in the Josephson array, of a few tens of $\mathrm{\mu A}$, is well below the current amplitude of the Shapiro steps and ensures a  perfect quantization of the QHRS (Appendix C.2). The PJVS and the QHRS are individually checked following the usual technical guidelines\cite{Behr2003,Delahaye2003}.

A simple connection of the PJVS to the QHRS would not allow realizing $U_{\mathrm J}/R_\mathrm{H}$ with the highest accuracy because of the large value of the two-wires series resistance (symbolized by $r$ in Fig.1c) caused by the connecting links. A multiple series connection of the QHRS\cite{Delahaye1993,Poirier2014}, a technique which exploits fundamental properties of the QHE, is implemented to reduce their effect. Each superconducting pad of the PJVS is connected to two QHRS terminals located along an equipotential edge of the Hall bar (Fig.2a). Due to the chirality of the Hall edge-states for the given magnetic field direction, $I_\mathrm{PQCS}$ essentially flows in the link of resistance ($r_1+r^{'}_1$) (typically $\sim 4~\Omega$). This gives rise to a voltage $(r_1+r^{'}_1)\times I_\mathrm{PQCS}$. Because of the edge equipotentiality and knowing that the two-terminal resistance is $R_\mathrm{H}$ in the QHE regime, a small current $i=(r_1+r^{'}_1)/R_\mathrm{H}\times I_\mathrm{PQCS}$ circulates this time, in the connection link probing the Hall voltage. This results in a small voltage, no more than $(r_1+r^{'}_1)r_2/R_\mathrm{H}\times I_\mathrm{PQCS}$, that adds to the Hall voltage $R_\mathrm{H}\times I_\mathrm{PQCS}$. This gives a relative correction to the quantized Hall resistance, $r/R_\mathrm{H}$, of $(r_1+r^{'}_1)r_2/R_\mathrm{H}^{2}$ (typically $9\times 10^{-8}$) much lower than $(r_1+r^{'}_1)/R_\mathrm{H}$ (typically $3\times 10^{-4}$) for a single connection. For some measurements, a third terminal was connected at top of the QHE cryostat (dotted line in Fig.2a) to further reduce the correction. In general, $I_\mathrm{PQCS}$ can be written as $\frac{U_\mathrm{J}}{R_\mathrm{H}}(1-\alpha)=2n_{\mathrm J}(K_\mathrm{J}R_\mathrm{K})^{-1}f_{\mathrm J}(1-\alpha)$, where $\alpha$ is a small relative correction that is calculated taking into account the resistance of all connections (Appendix D). In our experiments, this correction $\alpha$ of no more than $3\times10^{-7}$, is determined with an uncertainty of $u_{\alpha}=2.5\times10^{-9}$. The validity of this relationship giving $I_\mathrm{PQCS}$ assumes a perfect equipotentiality in the superconducting pads where voltage and current terminals of the QHRS are connected. The quantized current $I_\mathrm{PQCS}$ can be rewritten as $I_\mathrm{PQCS}\equiv n_{\mathrm J}ef_{\mathrm J}(1-\alpha)$, a form which is very similar to the expression of the current generated by the electron pumps, $n_{Q}ef_\mathrm{P}$. This reflects that both ways to realize the ampere, \emph{i.e.} from SET devices and from the application of Ohm's law to quantum standards (Fig.1a), are theoretically equivalent. It also points out that $I_\mathrm{PQCS}$ corresponds to the circulation of $n_\mathrm{J}$ elementary charges per cycle of the external frequency. Because $n_{\mathrm J}$ and $f_{\mathrm J}$ are orders of magnitude larger than $n_{\mathrm Q}$ and $f_\mathrm{P}$,the PQCS, and consequently the PQCG can generate higher currents than SET devices.
\begin{figure*}[t]
\includegraphics[width=17cm]{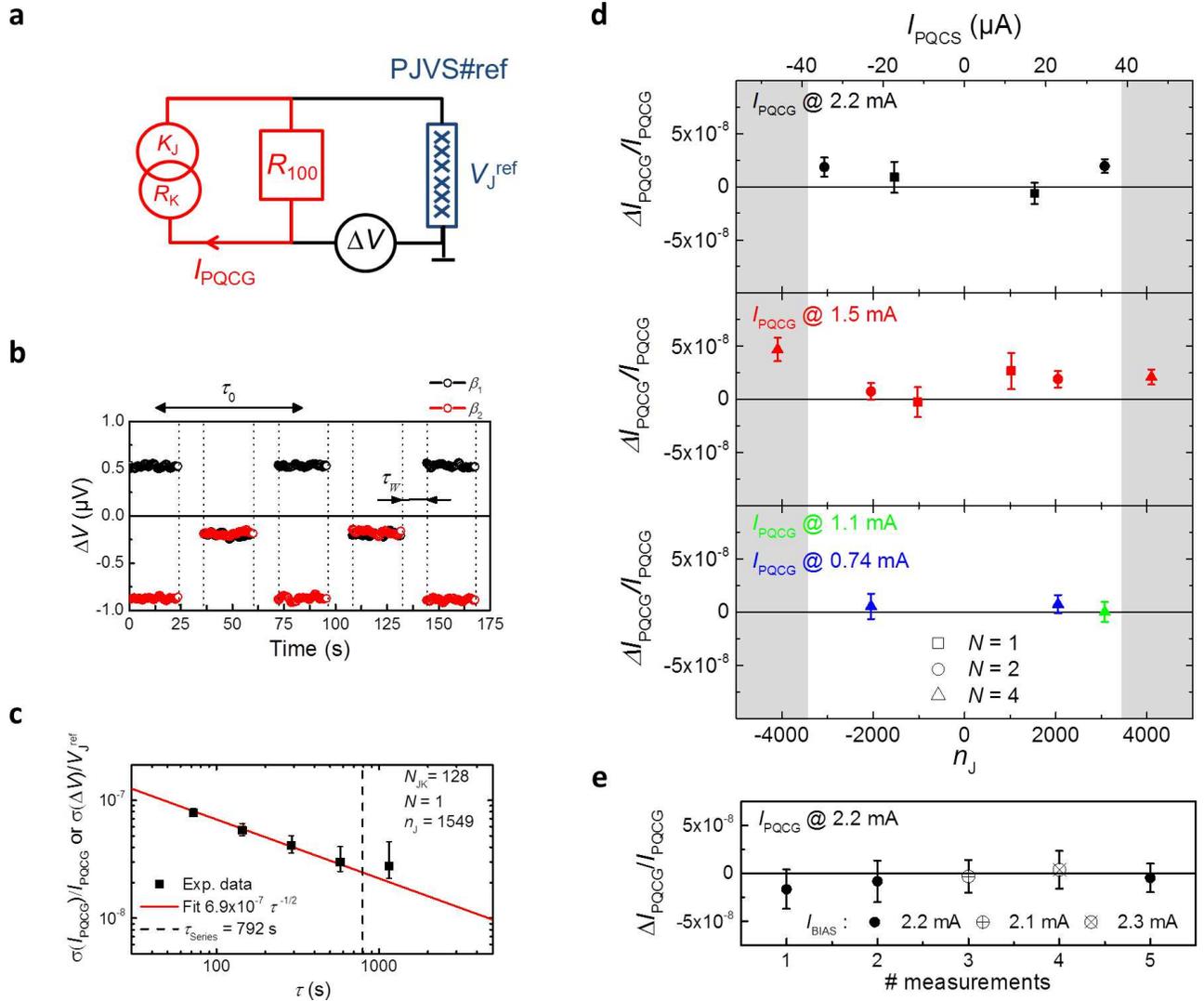}\hfill
\caption{\textbf{Quantization tests of the PQCG.} a) Scheme of the set-up for the accuracy measurements. The PQCG, symbolized by the current source linked to the product of constants $K_\mathrm{J}R_\mathrm{K}$, supplies a calibrated $100~\Omega$ resistor $R_\mathrm{100}$ and its voltage is compared to the voltage $V_\mathrm{J}^{ref}$ of PJVS\#ref using a battery-powered nanovoltmeter EM N31 (response time constant $\sim$ 2 s). b) Raw data of the voltage null detector for two on-off-on cycles for two settings $\beta_1$ (black dots) and $\beta_2$ (red dots) of the current divider. $\tau_0=72$ s is the duration of one on-off-on cycle. $\tau_w=12$ s is the waiting time before recording after the current switching. c) Relative Allan deviation calculated from a series of 49 on-off-on cycles of $I_\mathrm{PQCG}$ plotted as a function of time $\tau(s)$. A $\tau^{-1/2}$ fit shows a good agreement below $\tau=1000~$s and corresponds to a relative Allan deviation of $2.5\times10^{-8}$ for the duration of the time series used in this work $\tau_\mathrm{Series}=792~$s. d) $\Delta I_\mathrm{PQCG}/I_\mathrm{PQCG}$ as a function of $n_\mathrm{J}$ (or $I_\mathrm{PQCS}$) for several currents (both positive and negative) generated by the PQCG in the mA range, using three different values for $N$. Error bars are combined standard uncertainties (1 s.d.). e) Successive measurements of $\Delta I_\mathrm{PQCG}/I_\mathrm{PQCG}$ (over four hours) obtained, with $n_\mathrm{J}=3074$, $N=2$, $N_\mathrm{JK}=129$, for different bias currents demonstrating the reproducibility of the current generated at $2.2~\mathrm{mA}$. Error bars are combined standard uncertainties (1 s.d.).}\label{fig:Fig3}
\end{figure*}

The accuracy of the PQCG also relies on the detection and then the amplification with gain $G$ of the quantized current $I_\mathrm{PQCS}$ while keeping the accuracy. This is achieved using a cryogenic current comparator (CCC)\cite{Harvey1972}(see Appendix C.3) which is able to accurately compare currents with a relative uncertainty of a few $10^{-11}$. The CCC accuracy relies on Amp\`ere's theorem and the perfect diamagnetism of the superconductive toroidal shield (Meissner effect), in which several superconducting windings of variable number of turns are embedded. Its current sensitivity relies on a DC (direct current) superconducting quantum interference device (SQUID) detecting the flux generated by the screening current circulating on the shield (Fig.2a). More precisely, two windings of equal number of turns $N_\mathrm{JK}$ (128 or 129) are inserted in the connections to the QHRS on the low potential side of the circuit, \emph{i.e.} the grounded PJVS side, in order to detect the sum of the currents in the two windings, hence $I_\mathrm{PQCS}$. It is essential to cancel the leakage current that could alter the accurate equality of the total currents circulating in the QHRS and in these two windings. This is achieved by placing high and low potentials cables (high-insulation $R_\mathrm{L} > 1~\mathrm{T}\Omega$ resistance) connected to the QHRS inside two separated shields, that are then twisted together and connected to ground. By this way, direct leakage currents short-circuiting the QHRS, the most troublesome, are cancelled. Other leakage currents are redirected to ground. They lead to a relative error on the detected current, negligible, of no more than $(r_1+r^{'}_1)/R_\mathrm{L}\sim 4\times 10^{-12}$\cite{Poirier2014}. A third CCC winding, of number of turns $N$ (chosen between 1 and 4130), is connected to an external battery-powered and low-noise current source servo-controlled by the feedback voltage of the DC SQUID and which delivers the current $I_\mathrm{PQCG}$ so that the total ampere.turn in the CCC is zero ($N_\mathrm{JK}I_\mathrm{PQCS}-NI_\mathrm{PQCG}=0$). The current source is therefore locked to $I_\mathrm{PQCS}$ and generates a quantized current $I_\mathrm{PQCG}$ theoretically equal to $GI_\mathrm{PQCS}$, where $G=\frac{N_\mathrm{JK}}{N}$ spans two orders of magnitude above or below the unity gain. The CCC relying on a magnetic coupling between windings, it allows a high electrical-insulation between the PQCS and the external circuit connected to the devices under test that protects from an alteration of the PQCS quantization. $G$ is the main control parameter determining the range of the output current. The fine programmability of the quantized current $I_\mathrm{PQCG}$ can be achieved either by changing $n_{\mathrm J}$ or $f_\mathrm{J}$. Another option consists in tuning the CCC gain using a calibrated current divider that derives a fraction $\beta$ of the current in a fourth CCC winding ($N_\mathrm{Div}=16$ turns). In this case, the resulting CCC gain can be expressed as $G_{\beta}=\frac{N_\mathrm{JK}}{N+\beta N_\mathrm{Div}}$. $\beta$ can be varied over a range $\pm 5~\times10^{-5}$ and is determined with a standard uncertainty $u_{\beta}=0.5\times10^{-9}$. The accuracy in realizing the gain depends on the feedback electronics. To avoid significant error caused by the finite value of the amplifier gain of the SQUID electronics, the total ampere.turn value in the CCC is nominally strongly reduced by a fine tuning of the external current source using $V_\mathrm{out}$ as indicated in Fig.2a. Moreover, the gain of the feedback loop is set at the largest value possible that maintains the SQUID locked during on-off switchings of the current (controlled by the on-off switchings of $U_\mathrm{J}$). In these conditions, the relative quantization error of the PQCG, related to the finite open loop gain, is lower than $0.5\times10^{-9}$ (see Appendix E).

\subsection{\label{sec:level1}Noise, current uncertainty and stability}
The noise of the PQCG current, $S_I$, originates from $I_\mathrm{PQCS}$ and the gain $G_{\beta}$. It manifests itself in the flux detection by the SQUID of the CCC amplifier. $S_I$, can be expressed in relative value by $\frac{S_I}{I}(f)=\frac{1}{n_\mathrm{J}ef_\mathrm{J}}\frac{\gamma_\mathrm{CCC}}{N_\mathrm{JK}}S_\mathrm{\phi}(f)$, where $S_\mathrm{\phi}(f)$ is the flux noise amplitude density detected by the SQUID and $\gamma_\mathrm{CCC}=\mathrm{8~\mu A.turn/\phi_0}$ is the flux to ampere.turn sensitivity of the CCC. This expression shows that, the larger the number of Josephson junctions $n_\mathrm{J}$ and the number of turns $N_\mathrm{JK}$, the better the signal to noise. $S_\mathrm{\phi}$ results from the SQUID noise $S_\mathrm{SQUID}(f)$, the Johnson-Nyquist noise of the QHRS resistance and some external noise $S_\mathrm{ext}(f)$ captured by the measurement circuit. Note that the noise of the external current source servo-controlled by the SQUID is of no concern in the operation frequency bandwidth ($<$ 1 kHz) of the SQUID feedback. In these conditions, $S_\mathrm{\phi}(f)=\sqrt{S_\mathrm{SQUID}(f)^2 + \frac{4k_\mathrm{B}T}{R_\mathrm{H}}(\frac{N_\mathrm{JK}}{\gamma_\mathrm{CCC}})^2+ S_\mathrm{ext}(f)^2}$, where \emph{T}=1.3 K is the QHRS temperature. The flux noise density generated by the QHRS, of $\sim \mathrm{1~\mu\phi_0/Hz^{1/2}}$, is well below the base noise ($\mathrm{\sim 10~\mu\phi_0/Hz^{1/2}}$) measured by the SQUID operating in the bare CCC (Appendix C.3), as reported in Fig.2d (black curve).

Experimentally, the PQCG involves three quantum devices placed in independent cryogenic setups. The quantum devices are connected together using long shielded cables made of twisted pairs. Given the high-sensitivity of the SQUID to electromagnetic noise, achieving a stable and accurate operation of the PQCG was a challenge. To ensure the SQUID stability using $N_\mathrm{JK}$ as large as 129, it was necessary to connect a damping circuit to a fifth CCC winding ($N_\mathrm{D}=1600$) in order to avoid the amplification of the current noise in the PQCS loop at the resonance frequency of the CCC. It results that the noise spectrum $S_\phi(f)$ measured at the output of the SQUID presents a damped resonance at $\mathrm{1.6~kHz}$, much broader than the self-resonance of the bare CCC around 13 kHz (Fig.2d). Its amplitude is in good agreement with the Johnson-Nyquist noise emitted by the resistor $R_\mathrm{D}$ in the damping circuit and which is reported as the blue dashed line in Fig.2d (Appendix F). At frequencies between 0.1 Hz and 6 Hz, the noise level remains low and flat, with an amplitude of $\mathrm{\sim 20~\mu\phi_0/Hz^{1/2}}$ higher than the level in the bare CCC, indicating however that some external extra noise ($S_\mathrm{ext}(f)$) couples to the measurement circuit. Below 0.1 Hz, excess noise corresponding to a typical $1/f$ frequency dependence of the power spectral density $S_\mathrm{SQUID}^2$ was observed.

The noise $S_\phi(f)$ manifests itself in the current measurement by a relative standard uncertainty $u^\mathrm{A}_\mathrm{PQCG}$ (Type A evaluation) that can be evaluated by a statistical analysis of series of observations (Annexe B). The other significant contributions to the relative uncertainty of the PQCG, that are evaluated by non-statistical methods (Type B evaluation), comes from the cable correction ($u_{\alpha}$) and the current divider calibration ($u_{\beta}$) since components coming from frequency, QHRS, CCC, electronic feedback, and current leakage are negligible. These contributions result in a relative standard uncertainty $u^\mathrm{B}_\mathrm{PQCG}\simeq\sqrt{u_{\alpha}^2+(u_{\beta}\times N_\mathrm {Div}/N)^2}$ (Appendix G). In our experiments, it varies from only $2.5\times 10^{-9}$ (when the current divider is not used) up to $8.4\times 10^{-9}$ (for $N_\mathrm {Div}/N=16$). The combined standard uncertainty of $I_\mathrm{PQCG}$ can then be calculated from $\sqrt{{u^\mathrm{A}_\mathrm{PQCG}}^2+{u^\mathrm{B}_\mathrm{PQCG}}^2}$.

 Fig.2e reports on series of on-off switchings of the current $I_\mathrm{PQCG}$ at $1.1~\mathrm{mA}$ amplitude, as recorded by a digital ammeter. The current value was obtained for $n_\mathrm{J}=3073$, $N_\mathrm{JK}=129$ and $N=4$ ($G=129/4$). This figure demonstrates the capability of the PQCG to generate large currents, \emph{i.e.} in the milliampere range. It also reveals the low-noise level and the stability of the PQCG, notably characterized by the absence of SQUID unlocking at on-off switchings of the current.

\section{\label{sec:level1}Accuracy measurements of the PQCG}
The accuracy of the PQCG is determined by measuring the generated current $I_\mathrm{PQCG}$ and then comparing this measurement to its expected expression $2G_\beta n_\mathrm{J}(R_\mathrm{K}K_\mathrm{J})^{-1}f_\mathrm{J}(1-\alpha)$. Experimentally, the current $I_\mathrm{PQCG}$ generated by the PQCG, operated with PJVS\#A (Appendix C.1) is determined by measuring the voltage difference $\Delta V$ (using a EM N31 nanovoltmeter) between the voltage drop at the terminals of a very stable $100~\Omega$ resistance standard $R_\mathrm{100}$, calibrated in terms of $R_\mathrm{K}$ with an uncertainty of $2.5\times10^{-9}$, and the reference voltage $V_\mathrm{J}^\mathrm{ref}$ of a second PJVS (PJVS\#ref)(Appendix C.1), linked to $K_\mathrm{J}$, and operated synchronously at the same frequency $f_\mathrm{J}$ (Fig.3a). For experimental convenience, the frequencies of both PJVS were kept constant while the voltage balance on the null detector has been done by selecting appropriate number of Josephson junctions for both PJVS and by fine tuning the CCC gain $G_{\beta}$. Then, the experimental procedure consists in finding the fraction $\beta_0$ corresponding to equilibrium, \emph{i.e.} for $I_\mathrm{PQCG} = V_\mathrm{J}^\mathrm{ref}/R_\mathrm{100}$. This means that $I_\mathrm{PQCG}$ is measured through an identification with the reference quantized current, $V_\mathrm{J}^\mathrm{ref}/R_\mathrm{100}$, itself perfectly known in terms of $(R_\mathrm{K}K_\mathrm{J})^{-1}$. The accuracy of the PQCG is then expressed by the relative deviation $\Delta I_\mathrm{PQCG}/I_\mathrm{PQCG} = (I_\mathrm{PQCG}-G_{{\beta}_{0}} I_{\mathrm{PQCS}})/I_\mathrm{PQCG}$ between the measured current $I_\mathrm{PQCG}$ (=$V_\mathrm{J}^\mathrm{ref}/R_\mathrm{100}$) and the current $G_{{\beta}_{0}} I_{\mathrm{PQCS}}$ calculated from $\beta_0$.
In practice, $\beta_0$ is determined from the successive measurements of two small voltages $\Delta V_1$ and $\Delta V_2$, obtained for two settings $\beta_{1}$ and $\beta_{2}$ chosen above and below $\beta_0$ respectively, and the linear relationship $\beta_0=\beta_1+(\beta_2-\beta_1)\times\Delta V_1/(|\Delta V_1|+|\Delta V_2|)$ (Appendix H). Moreover, series of on-off-on cycles illustrated in Fig.3b are used to subtract voltage offsets and truncate the 1/\emph{f} noise of the CCC at the repetition frequency $1/\tau_0$ of the cycles. The noise of the measurements is analyzed with the help of the Allan deviation\cite{Witt2005} (Appendix I), which allows to distinguish between the different types of noise according to the exponent of its power dependence with time. The efficiency of the 1/\emph{f} noise rejection procedure is demonstrated by Fig.3c which reports the typical time dependence of the relative Allan deviation of the voltage $\sigma (\Delta V)/V^\mathrm{ref}_\mathrm{J}$ (or equivalently of the current $\sigma(I_\mathrm{PQCG})/I_\mathrm{PQCG}$). The $\tau^{-1/2}$ behavior is typical of a white noise regime and legitimates the calculation of experimental standard deviation of the mean for the 11-cycles time series (792 s duration) to evaluate the standard uncertainties (Type A evaluation) $u_\mathrm{\Delta V_1}$ and $u_\mathrm{\Delta V_2}$, of $\Delta V_1$ and $\Delta V_2$ respectively. These uncertainties, of no more than about $2.5\times10^{-8}$ of $V^\mathrm{ref}_\mathrm{J}$ for $n_\mathrm{J}=1549$, are then combined with $u_{\beta_1}$ and $u_{\beta_2}$ (evaluated by Type B methods) to determine the standard uncertainty $u_{\beta_0}$ of $\beta_0$ (Appendix H). The latter is the main contribution to the measurement uncertainty of $\Delta I_\mathrm{PQCG}/I_\mathrm{PQCG}$ (Appendix J). A $10^{-8}$ uncertainty is typically achievable for $n_\mathrm{J}=3072$ and an experiment duration of 1600 s.

Fig.3d shows the relative deviation $\Delta I_\mathrm{PQCG}/I_\mathrm{PQCG}$ as a function of the number $n_\mathrm{J}(\propto I_\mathrm{PQCS})$ of biased Josephson junctions at four different amplitudes of $I_\mathrm{PQCG}$ in the mA range. Note that maintaining the output current for different $n_\mathrm{J}$ requires varying the number of turns $N$ in order to keep the ratio $n_\mathrm{J}/N$ constant.
Each reported data represents the arithmetic mean value of measurements carried out at different moments. The data show no significant deviation within combined (including Type A and Type B uncertainty contributions) relative standard uncertainties (1 s.d.) of less than $2\times10^{-8}$ in relative value whatever the value of $n_\mathrm{J}$, except for $n_\mathrm{J}=\pm 4098$ and $I_\mathrm{PQCG}$ at $1.5~\mathrm{mA}$.

Indeed, for values of $n_\mathrm{J} > 3074$, \emph{i.e.} a current $I_\mathrm{PQCS} > \mathrm{35~\mu A}$, one observes an increased dispersion of the experimental data for $I_\mathrm{PQCG}$ with significant deviations from theoretical values. These deviations are not clearly understood at the present time. The usual individual quantization tests of both PJVS\#A and PJVS\#ref and of the QHRS have confirmed that these deviations were not caused by a lack of voltage quantization of the voltage steps nor of the quantum Hall resistance plateau. Nevertheless, the discrepancies increased with time and were sensitive to room temperature cycling of PJVS\#A. An alteration of the perfect equipotentiality in the superconducting pads of PJVS\#A caused by the circulation of the $I_\mathrm{PQCS}$ current might be considered but will need further investigations.

For $n_\mathrm{J}\leq 3074$, such accuracy deterioration have also been observed occasionally after a long period of operation (several hours or a day), the phenomenon being less pronounced and more rare at low $n_\mathrm{J}$. However, the quantization of the current $I_\mathrm{PQCG}$ was always fully restored by a room temperature cycling of PJVS\#A, the Josephson array through which $I_\mathrm{PQCS}$ circulates. To illustrate the time reproducibility of the current quantization at low $n_\mathrm{J}$ after a cycling, Fig.3e reports successive measurements of $\Delta I_\mathrm{PQCG}/I_\mathrm{PQCG}$, here carried out over four hours. All results are close to zero within a relative uncertainty of $10^{-8}$ for $I_\mathrm{PQCG}$  at $2.2~$mA and $n_\mathrm{J}=3074$ ($N_\mathrm{JK}=129$). This figure also demonstrates the independence of $\Delta I_\mathrm{PQCG}/I_\mathrm{PQCG}$ as a function of $I_{Bias}$, \emph{i.e.} the current used to bias the junctions of PJVS\#A, over $\mathrm{\pm 0.1 mA}$ from the voltage step center at 2.2 mA. This property is a necessary quantization criterion.

Following these considerations, Fig.1d was elaborated by averaging, for each output current value, the data reported in Fig.3d obtained at different $n_\mathrm{J}\leq 3074$. For each current value from $\pm 0.7$ to $\pm 2.2 \mathrm{mA}$, no significant relative deviation $\Delta I_\mathrm{PQCG}/I_\mathrm{PQCG}$ is observed considering the combined measurement uncertainty of $1\times 10^{-8}$ (1 s.d.). Moreover, the experimental standard deviation of all data points over the whole range amounts only to $8\times10^{-9}$. Finally, the weighted mean of $\Delta I_\mathrm{PQCG}/I_\mathrm{PQCG}$ is equal to $(6\pm6)\times10^{-9}$. These results demonstrate the quantization accuracy of the PQCG in terms of $(K_\mathrm{J}\times R_\mathrm{K})^{-1}\equiv e/2$, within a $10^{-8}$ relative uncertainty in the mA range.

The current value $I_\mathrm{PQCG}$ results not only from the $I_\mathrm{PQCS}$ value but also from the amplification gain $G=N_\mathrm{JK}/N$ which is highly-accurate and can span two orders of magnitude above or below the unity gain ($N$ is between 1 and 4130, $N_\mathrm{JK}$ is fixed). Therefore, $I_\mathrm{PQCG}$ is quantized with the same accuracy over the wide range of current values accessible by changing $G$, while $I_\mathrm{PQCS}$ remains below $\mathrm{35~\mu A}$, \emph{i.e.} $n_\mathrm{J}\leq 3074$. This upper limit for $I_\mathrm{PQCS}$ is close to the current value used to bias the GaAs/AlGaAs-based QHRS in optimized resistance calibration, thus does not restrict the PQCG use. Moreover, the relative current density noise $S_I/I$ does not depend on $G$ ($N_\mathrm{JK}$ is fixed), but only on $n_\mathrm{J}$ (\emph{i.e.} $I_\mathrm{PQCS}$). Considering $n_\mathrm{J}=3074$ that gives the best signal to noise ratio, one concludes that the PQCG can accurately generate currents with a combined relative measurement uncertainty of $10^{-8}$ in the whole range from $\mathrm{1~\mu A}$ up to 10 mA, as illustrated in Fig.1b.

\section{\label{sec:level1}Using the PQCG for current traceability}
\begin{figure}[t]
\includegraphics[width=7.3cm]{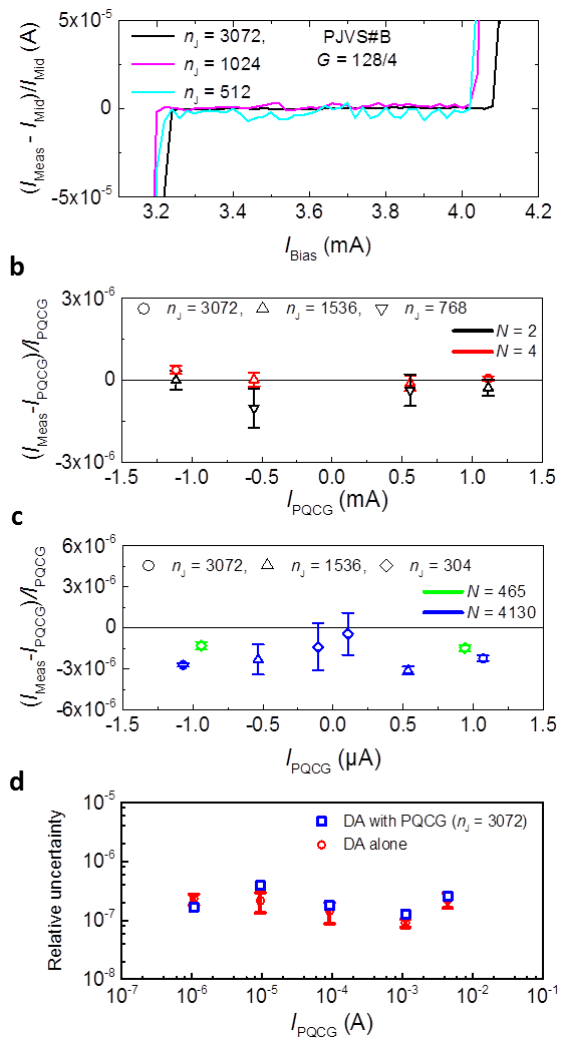}\hfill
\caption{\textbf{Calibration of a digital ammeter using the PQCG and identification of quantization criteria.} a) Relative deviation of $I_\mathrm{Meas}$ (the current generated by the PQCG and measured by the DA (HP3458A)), to $I_\mathrm{Mid}$ (the current measured at the center of the step) as a function of the biasing current $I_\mathrm{Bias}$ for $n_\mathrm{J}=3072$, 1024 and 512 junctions of PJVS\#B. The independence of the output current as a function of $I_\mathrm{Bias}$ is a first quantization criterion. b) Relative deviation of $I_\mathrm{Meas}$ to $I_\mathrm{PQCG}$ as a function of $I_\mathrm{PQCG}$ in the mA range and c) in the $\mathrm{\mu A}$ range. The agreement of the measurements performed using $n_\mathrm{J}$ and $n_\mathrm{J}/2$ is a second quantization criterion. d) Relative uncertainty of the measured current (blue square points) and of the DA (open red dot) as a function of $I_\mathrm{PQCG}$ over four decades of current. The DA noise dominates that of the PQCG. Error bars are standard uncertainties (1 s.d.).}\label{fig:Fig4}
\end{figure}
These high-accuracy measurements have validated the PQCG as a quantum current standard. However it remains important to demonstrate that the PQCG, once checked using quick quantization criteria, can be used to calibrate a commercial digital ammeter (DA) over several current ranges. This has been realized by replacing the load in Fig.2a by a precision DA (a HP3458A multimeter - Appendix K) with the low potential input connected to ground. Connecting the DA directly to the output of the PQCG is an extra challenge due to the sensitivity of quantum devices (in particular the SQUID) to the environmental noise. It has been possible at the expense of shunting the differential input by a 100 nF highly-insulated Polytetrafluoroethylene (PTFE) capacitance that short-circuits some digital noise generated by the DA. Prior to the calibration, the adjustment procedure recommended by the manufacturer has been followed.

We have then performed current measurements in the DA ranges from 10 mA down to $\mathrm{1~\mu A}$ by changing the gain $G$ ($N_{\mathrm{JK}}=$128 and $N$ spanning from 1 to 4130) while using the highest number of Josephson junctions possible when $n_\mathrm{J}$ had to be reduced below 3072, in order to optimize the signal to noise ratio. Two quantization criteria have been identified. The first one is the independence of the output current as a function of the current biasing the Josephson array. Fig.4a shows three quantized current steps which are flat within a few $10^{-6}$ at $I_\mathrm{Meas} \sim \mathrm{1.1~mA}$, $\mathrm{0.37~mA}$ and $\mathrm{0.18~mA}$. These  currents steps were obtained by varying the bias current of Josephson array segments of PJVS\#B (Appendix C.1) containing 3072, 1024 and 512 Josephson junctions respectively. Note that the operating current margins are the same as those of the corresponding voltage steps. The second quantization criterion is the independence of the calibration results obtained with different values of $n_\mathrm{J}$ for the same output current. This is illustrated in Fig.4b that reports the relative deviation, obtained from eight on-off-on cycles over about 15 minutes, between the measured current $I_\mathrm{Meas}$ and the quantized current $I_\mathrm{PQCG}=GI_\mathrm{PQCS}$ (see Appendix K for calculation of $I_\mathrm{PQCG}$ values). In the 1 mA range, the same currents $I_\mathrm{PQCG}$ have been generated by biasing both $n_\mathrm{J}$ ($N=4$) and $n_\mathrm{J}/2$ ($N=2$) Josephson junctions. The relative deviations are in agreement within the measurement uncertainties (see Appendix K) which confirms that the current generated by the PQCG is independent of the value of $I_\mathrm{PQCS}$, provided that it is lower than $\mathrm{35~\mu A}$ ($n_\mathrm{J}\leq 3074$).

The lowest uncertainty measurements of Fig.4b show that the DA is accurate and linear within a relative uncertainty of $5\times 10^{-7}$, which is better than the manufacturer specifications (see Appendix K). The same measurements were performed on the $\mathrm{1~\mu A}$ range. The results show a significant deviation from $I_\mathrm{PQCG}$ of about $3\times 10^{-6}$ and a higher dispersion of the data points which are due to the accuracy limitation and bigger instability in this range, as can be deduced from the manufacturer specifications. It is important to note that however a relative uncertainty of $\sim 2\times 10^{-7}$ is achieved for measurements at the top of both ranges presented in Fig.4b and 4c. Note that this is the case for all ranges studied in this paper as it is demonstrated in Fig.4d, which reports the relative uncertainties of the measurements performed with $n_\mathrm{J}=3072$ and by varying $N$ from 1 up to 4130. Fig.4d shows that the current noise of the PQCG is independent of $N$ and that the measurements uncertainty is dominated by the noise of the DA (red data points) (Appendix K). This could be expected from the DA specifications and the low-noise of the PQCG demonstrated in Section III (about $10^{-8}$ uncertainty for similar measurement time and $n_\mathrm{J}$ value).  More surprisingly, for $n_\mathrm{J}\leq 1536$, the PQCG noise appears overcoming the one of the DA in the 1 mA range since dividing $n_\mathrm{J}$ by a factor of two for a given current value, doubles the measurement uncertainties (a few $10^{-7}$) reported in Fig.4b. This is not the case in the $1~\mu$A range while decreasing $n_\mathrm{J}$ by a factor of ten (Fig.4c). This can be explained by the noise spectrum of the PQCG (Fig. 2d.) and the bandwidth of the measurements which depends on the current ranges, due to the presence of the 100 nF capacitance forming a low-pass filter with the input resistance of the DA, with cutoff frequency decreasing at lower-current ranges (Appendix K). This filter prevents the medium-frequency noise from the damping circuit to overcome the noise of the DA for the low-current ranges. Hence, to fully benefit from the low-noise at low frequencies of the PQCG (as demonstrated in the PQCG accuracy measurements) when calibrating the DA in all current ranges, improvements of the filtering will be carried on in the future. Simultaneously, cooling down the damping resistor $R_\mathrm{D}$ responsible for the Johnson-Nyquist noise will decrease significantly the PQCG noise at medium frequencies and will also be implemented in the future.
\section{\label{sec:level1}Discussion}
Returning towards Ohm's law, which is the basis for the definition of the resistance unit, we developed a quantum current standard from the quantum Josephson voltage and Hall resistance standards that are combined in an original quantum circuit, with the aim of universality, accuracy and simplicity\cite{Kibble2010}. The programmable quantum current generator (PQCG) reported here, is able to generate currents from $\mathrm{1~\mu A}$ up to 5 mA values, that are quantized in terms of $(K_\mathrm{J}\times R_\mathrm{K})^{-1}$ with a $10^{-8}$ relative standard uncertainty. This universal and versatile quantum current standard improves the accuracy of the current sources of two orders of magnitude compared to CMCs. It opens the way to a renewed metrology of the electrical current, that will also rely on the development of more stable current transfer standards. As a first proof of its impact, we showed that the PQCG, after identifying quantization criteria, can be used to calibrate efficiently a digital ammeter with measurement uncertainties only limited by the device under test.

Many improvements and extensions of the PQCG can be further considered. First, one can expect a noise reduction, typically by a factor of ten, by increasing the number of ampere.turns in the CCC, which can be achieved by a larger number of turns $N_\mathrm{JK}$ of the detection windings (up to 1600) and also by a higher current $I_\mathrm{PQCS}$ (for example, by increasing $n_\mathrm{J}$ values while preserving the accuracy). In any case, the damping circuit of the CCC resonances should be adapted and refined. In this work, the multiple connection of the QHRS was successfully implemented using different cable configurations (Appendix D), thanks to a correct evaluation of the cable correction in the PQCG expression. Nevertheless, the implementation of a complete triple connection of the QHRS will make the cable correction negligible and therefore simplify further the PQCG use. From all these improvements, the target uncertainty of $10^{-9}$ should be reached. Beyond, the availability of graphene-based quantum resistance standards operating in relaxed experimental conditions\cite{Ribeiro2015} should allow the implementation of the quantum voltage, resistance and current standards, as well as their combination, in a unique compact cryogen-free setup. This would constitute a major step towards the realization of a universal and practical quantum generator/multimeter.

More generally, the principle of using the PQCS as a reference for building the PQCG is seminal and can be exploited for other experiments or instruments\cite{Poirier2014}. For instance, a quantum current generator working in the AC regime can be developed using pulse-driven Josephson standards\cite{Benz2015,Kieler2015}, AC QHRS\cite{Ahlers2009} and current transformers\cite{Kibble1984}. This is a perspective that cannot be considered with single-electron sources in the present state-of-the-art. Very accurate and sensitive comparisons of quantum Hall resistances can be performed by opposing, by means of the CCC, the PQCS currents obtained from two different QHRS polarized by the same Josephson voltage reference. This novel comparison technique could be used to test the universality of the QHE from the integer to the fractional regime. Finally, a quantum ammeter\cite{Poirier2014} can be realized by directly comparing the current delivered by an external source to the PQCS using the CCC.

More fundamentally, the PQCG can implement the planned new definition of the ampere with the target uncertainty of $10^{-8}$ since it is linked to the elementary charge \emph{e}. Our work therefore provides an essential piece to the revised SI founded on constants of physics. This achievement will rely on the adoption of the fundamental relationships for the quantum Hall and Josephson effects in the future SI, which is also necessary for the realization of the kilogram from the Planck constant \emph{h} using the watt balance experiment\cite{Kibble1976}. Indeed, it relies on comparing the mechanical power with the electrical power, calibrated itself from the quantum voltage and resistance standards. In this context, the closure of the metrological triangle\cite{Likharev1985,Scherer2012} which consists in comparing the ampere realizations in terms of both $(K_\mathrm{J}\times R_\mathrm{K})^{-1}$ and $Q$, is an important and long awaited experiment. It indeed leads to the direct measurement of the product $R_\mathrm{K}\times K_\mathrm{J}\times Q$, theoretically equal to 2. Validating this equality with a measurement uncertainty down to $10^{-8}$, would strengthen the confidence in the description, in terms of \emph{h} and \emph{e} only, of the constants involved in the three solid-state quantum physics phenomena. In this perspective, the PQCG used as a reference and the quantum ammeter built on the PQCS, could be used to accurately measure SET-based current sources\cite{Giblin2012,Stein2015,Jehl2013} in terms of $(R_\mathrm{K}\times K_\mathrm{J})^{-1}$, in a more direct way than in previous experiments\cite{Devoille2012}.
\section*{\label{sec:level1}Acknowledments}
We wish to acknowledge, F. Lafont for fruitful discussions in the early stage of the experiment design, D. Est\`eve and Y. De Wilde for critical reading and comments, R. Behr for useful advice in using PTB Josephson arrays. This research was partly supported by the project 15SIB08 e-SI-Amp funded by the European Metrology Programme for Innovation and Research (EMPIR). The EMPIR initiative is co-funded by the European Union's Horizon 2020 research and initiative programme and the EMPIR participating states.\\
\section{\label{sec:level1}Appendix: methods}
\subsection{\label{sec:level1}volt, ohm and ampere representations}
 The uncertainties of $4\times10^{-7}$\cite{KJ} and $1\times10^{-7}$\cite{RK} on the determinations of $K_\mathrm{J}$ and $R_\mathrm{K}$ in SI units respectively does not allow benefitting from the high-reproducibility of the Josephson and quantum Hall effects for the traceability of the volt and the ohm. To overcome this limitation conventional values for $K_\mathrm{J}$ and $R_\mathrm{K}$ were recommended in 1990 by the Comit\'e International des Poids et Mesures (CIPM)\cite{RK} for the traceability of the voltage and the resistance in calibration certificates based on the implementation of these quantum effects. These constants are exact and given by $K_\mathrm{J-90}=483597.9~\mathrm{GHz/V}$ and $R_\mathrm{K-90}=25812.807~\Omega$. They are related to $K_\mathrm{J}$ and $R_\mathrm{K}$ through $K_\mathrm{J}=K_\mathrm{J-90}(1\pm4\times10^{-7})$ and $R_\mathrm{K}=R_\mathrm{K-90}(1\pm1\times10^{-7})$. The voltage and the resistance traceable to $K_\mathrm{J-90}$ and $R_\mathrm{K-90}$ give representations of the volt and the ohm, and not realization of the unit volt and the unit ohm (SI). It results that the current realized by application of Ohm's law from the representations of the volt and the ohm based on $K_\mathrm{J-90}$ and $R_\mathrm{K-90}$ gives a representation of the ampere, not spoiled by the uncertainties of $K_\mathrm{J}$ and $R_\mathrm{K}$.

 The new SI, that notably adopts exact values for \emph{h} and \emph{e} and a new definition of the ampere from \emph{e}, aims at solving this problem. If the relationships $K_\mathrm{J}=h/2e$ and $R_\mathrm{K}=h/e^2$ are adopted, the constants involved in the Josephson effect and the quantum Hall effect will no more have uncertainties. As a consequence, the Josephson voltage standard and the quantum Hall resistance standard will become SI realizations of the volt and ohm. The combination of these two quantum effects, as proposed in this paper, will lead to a SI realization of the ampere.
\subsection{\label{sec:level1}Uncertainty vocabulary}
This section reports the definitions from the GUM (Guide to the expression of uncertainty in measurement)\cite{GUM} of the metrological terms used in the main text.\\
\textbf{Uncertainty (of measurement):} parameter, associated with the result of a measurement, that characterizes the dispersion of the values that
could reasonably be attributed to the measurand.\\
\textbf{Standard uncertainty:} uncertainty of the result of a measurement expressed as a standard deviation.\\
\textbf{Type A evaluation (of uncertainty):} method of evaluation of uncertainty by the statistical analysis of series of observations.\\
\textbf{Type B evaluation (of uncertainty):} method of evaluation of uncertainty by means other than the statistical analysis of series of observations.\\
\textbf{Combined standard uncertainty:} standard uncertainty of the result of a measurement when that result is obtained from the values of a number
of other quantities, equal to the positive square root of a sum of terms, the terms being the variances or covariances of these other quantities weighted according to how the measurement result varies with changes in these quantities.
\subsection{\label{sec:level1}Quantum devices}
\subsubsection{\label{sec:level1}PJVS devices}
The three PJVS used in this work (PJVS\#A, PJVS\#B and PJVS\#ref) are based on 1 V $\mathrm{Nb/Al/AlO_x/Al/AlO_x/Al/Nb}$ Josephson junction series arrays fabricated at the Physikalisch-Technische Bundesanstalt (PTB) \cite{Krasnopolin2002,Mueller2007}.
PJVS\#A (data shown in Fig.2 and Fig.3) and PJVS\#B (data shown in Fig.4) were used in the PQCS. PJVS\#ref was used for the opposition voltage $V_\mathrm{J}^\mathrm{ref}$ in the accuracy measurements of the PQCG in the Fig.3.
The Josephson arrays are subdivided in 14 smaller array segments. PJVS\#A follows a sequence 256/512/3072/2048/1024/128/1/1/2/4/8/16/32/64, PJVS\#B and PJVS\#ref follow the sequence 4096/2048/1024/512/256/128/1/1/2/4/8/16/32/64. For PJVS\#ref, only few segments were used corresponding to a maximum number of junctions of 1920. The Table below sums up the characteristics of the arrays, where $I_{C} $ is the critical current, $\Delta I_{Bias}$ the current amplitude of the Shapiro steps, $P_{RF}$ is the microwave power applied at the array.
\begin{table}[htbp]
\begin{center}
\begin{tabular}{|l|c|c|c|c|}
  \hline
    & PJVS\#A & PJVS\#B & PJVS\#ref \\ \hline
  $I_{C}~(\mathrm{mA})$ & 1.4 & 1.6 & 3.5 \\
  $\Delta I_{Bias}~(\mathrm{mA})~@n=0$ & 0.6 & 1.1 & 3.1 \\
  $\Delta I_{Bias}~(\mathrm{mA})~@n=1$ & 0.6 & 0.8 & 1.0 \\
  $I_{Mid}~(\mathrm{mA})$ & 2.4 & 3.65 & 4.4 \\
  $n_\mathrm{J}$ & 7168 & 8191 & 1920 \\
  $f_\mathrm{J}~(\mathrm{GHz})$ & 70 & 70.111 & 70 \\
  $P_{RF}~(\mathrm{mW})$ & 30 & 10 & 65 \\
  \hline
\end{tabular}
\caption{PJVS devices characteritics\label{PJVS devices}}
\end{center}
\end{table}
Note that a Josephson junction is missing in PJVS\#ref but acts as a perfect short circuit \cite{Mueller2007}. The \emph{I}-\emph{V} characteristics for the total number of junctions of the arrays have been systematically checked prior and after the current measurements. The microwave synthesizer is locked to a 10 MHz reference, delivered by a GPS Rubidium frequency standard.
\subsubsection{\label{sec:level1}QHRS device}
The Hall resistance standard, based on a eight-terminals Hall bar made of GaAs/AlGaAs semiconductor heterostructure (LEP514), was produced at the Laboratoire Electronique de Philips\cite{Piquemal1993}. Fig.2c reports the Hall resistance $R_{\mathrm H}$ and the longitudinal resistance per square $R_{xx}$ measured as a function of \emph{B}. The metrological quality of the sample was checked following the technical guidelines\cite{Delahaye2003}. At \emph{B}=10.8 T, $T=1.3~\mathrm{K}$ and for currents below $\mathrm{60~\mu A}$, $R_{\mathrm H}$ is perfectly quantized at $R_\mathrm{K}/2$ within a relative uncertainty of $1\times10^{-10}$ and the two-dimensional electron gas is dissipation-less ($R_{xx}\leq 10~\mathrm{\mu\Omega}$)\cite{Ribeiro2015}. The resistance of the eight contacts is lower than $0.1~\Omega$.
\subsubsection{\label{sec:level1}CCC device}
The cryogenic adder-amplifier is based on a cryogenic current comparator (CCC) usually used in a bridge performing accurate resistance comparisons. More precisely, it is made of fifteen windings with the following numbers of turns: 1, 1, 2, 2, 16, 16, 32, 64, 128, 160, 160, 1600, 1600, 2065 and 2065. It is equipped with a Quantum Design Inc. DC SQUID having a $\mathrm{3~\mu \phi_0/Hz^{1/2}}$ base white noise\cite{Lafont2015}. Fig.2d reports the noise spectral density $S_\phi$ measured at the output of the SQUID as a function of the frequency for the CCC alone (no winding connected). The bottom white noise level is around $\mathrm{10~\mu \phi_0/Hz^{1/2}}$ indicating that some external noise is captured. Considering the CCC ampere.turn gain of $G_\mathrm{CCC}=\mathrm{8~\mu A.turn/\phi_0}$, this corresponds to a $\mathrm{80~pA.turn/Hz^{1/2}}$ current sensitivity. At frequencies above 10 kHz, intrinsic electrical resonances of the CCC due to its high inductance are observable. From a few kilohertz down to 6 Hz, Fig.2d displays peaks which are caused by mechanical and acoustic resonances. At lower frequencies, $S_\phi$ features white noise down to 0.1 Hz and for even lower frequencies it rises according to $1/f^{1/2}$ due to the 1/\emph{f} SQUID noise.
\subsection{\label{sec:level1}Cable corrections $\alpha$}
 The use of multiple series connections to the QHRS reduces drastically the positive correction to the quantized Hall resistance caused by the resistance of these connections\cite{Delahaye1993,Poirier2014}. It results in a negative relative correction $\alpha$ added to the quantized current $n_{\mathrm J}ef_{\mathrm J}$ leading to $I_\mathrm{PQCS}=n_{\mathrm J}ef_{\mathrm J}(1-\alpha)$. Considering the link resistances $r_{1}$, $r^{'}_{1}$, $r_{2}$, $r_{3}$, $r^{'}_{3}$, and $r_{4}$, as indicated in Fig.2a, one calculates, using a Ricketts and Kemeny model\cite{Ricketts1988} of the Hall bar, $\alpha=\frac{(r_{1}+r^{'}_{1})r_{2}}{R_\mathrm{H}^2} + \frac{(r_{3}+r^{'}_{3})r_{4}}{R_\mathrm{H}^2}$ for the double connection scheme and $\alpha$ is reduced to $\alpha=\frac{(r_{1})r_{2}}{R_\mathrm{H}^2} + \frac{(r_{3})r_{4}}{R_\mathrm{H}^2}$ if a third terminal is connected at the top of the QHE setup. For the double connection scheme, we determined $\alpha=1.94\times 10^{-7}$ for $N_\mathrm{JK}=128$ and $\alpha=2.99\times 10^{-7}$ for $N_\mathrm{JK}=129$. With a third terminal connected $\alpha=1.16\times 10^{-7}$ for $N_\mathrm{JK}=128$ and $\alpha=2.20\times 10^{-7}$ for $N_\mathrm{JK}=129$. The total resistance of each connection being measured with a $50~m\Omega$ uncertainty, $\alpha$ is determined with a $u_{\alpha}=2.5\times10^{-9}$ relative standard uncertainty.
 \subsection{\label{sec:level1}Impact of the feedback settings on the PQCG accuracy}
To maintain the PQCG stable, even during the on-off switching of the current, the control voltage $V_\mathrm{out}$ was well adjusted so that the nominal ampere.turn value remained close to zero and the feedback gain of the SQUID was reduced so that the closed loop gain (CLG) is increased from $0.75~\mathrm{V/\phi_0}$ (the value in internal feedback mode operation) to $4.2~\mathrm{V/\phi_0}$ for $N$ values from 1 up to 465. For $N=4130$, the CLG was further increased up to $8.4~\mathrm{V/\phi_0}$. One can expect a quantization error of the PQCG resulting from the finite amplification gain (FAG) of the SQUID electronics (in $\mathrm{V/\phi_0}$) that leads to a non-zero ampere.turn value in the CCC. The relative current error is given by $\Delta I_\mathrm{PQCG}/I_\mathrm{PQCG}=-\mathrm{CLG/(CLG+FAG)\Delta^\mathrm{adj}}I_\mathrm{PQCG}/I_\mathrm{PQCG}$, where $\Delta^{\mathrm{adj}} I_\mathrm{PQCG}=G_{\beta}I_\mathrm{PQCS}-I^{\mathrm{adj}}_\mathrm{PQCG}$ is the deviation between the target quantized current $G_{\beta}I_\mathrm{PQCS}$ and the adjustment current $I^{\mathrm{adj}}_\mathrm{PQCG}$. $\Delta^{\mathrm{adj}} I_\mathrm{PQCG}$ can be determined from the SQUID output voltage equal to $N\Delta^{\mathrm{adj}} I_\mathrm{PQCG}\times\mathrm{CLG}/\gamma_\mathrm{CCC}$. The SQUID electronics amplifier being based on an integrator, it results that $\mathrm{FAG}\propto 1/f$, where $f$ is the measurement frequency. The error caused by the FAG is therefore nulled in the direct current limit (DC). To estimate the error on the quantized current generated by the PQCG in normal operation, we performed accuracy measurements using the usual on-off switchings frequency while intentionally shifting $I^{\mathrm{adj}}_\mathrm{PQCG}$ from adjustment. It turns out that increasing $\Delta^{\mathrm{adj}} I_\mathrm{PQCG}/I_\mathrm{PQCG}$ up to $10^{-3}$ leads to relative errors $\Delta I_\mathrm{PQCG}/I_\mathrm{PQCG}$ amounting to $(2.3 \pm 1.3)\times10^{-8}$. This corresponds to $\mathrm{FAG}\sim 4.3\times10^{4}\times \mathrm{CLG}\sim 1.8\times10^{5}~\mathrm{V/\phi_0}$. $\Delta^{\mathrm{adj}} I_\mathrm{PQCG}/I_\mathrm{PQCG}$ being maintained below $2\times10^{-5}$ in accurate operation of the PQCG, we deduce a relative error on the current generated $I_\mathrm{PQCG}$ of $(4.6\pm2.6)\times10^{-10}$, i.e. lower than $10^{-9}$.
\subsection{\label{sec:level1}Noise generated by the Damping circuit}
The Johnson-Nyquist noise of the $R_\mathrm{D}=1.1~\mathrm{k\Omega}$ resistance of the damping circuit placed at room temperature $T_\mathrm{D}=300~\mathrm{K}$ leads to the circulation of a noise current of density $\delta i(f)=\frac{jC_\mathrm{D}2\pi f\sqrt{4k_\mathrm{B}T_\mathrm{D}R_\mathrm{D}}}{1+R_\mathrm{D}jC_\mathrm{D}2\pi f-L_\mathrm{D}C_\mathrm{D}(2\pi f)^2}$, where $C_\mathrm{D}=100~\mathrm{nF}$ is the capacitance of the damping circuit and $L_\mathrm{D}=70~\mathrm{mH}$ is the inductance of the winding of $N_\mathrm{D}=1600$ number of turns. This results in a flux noise density of modulus $|S^\mathrm{D}_{\phi}(f)|=\frac{N_\mathrm{D}C_\mathrm{D}2\pi f\sqrt{4k_\mathrm{B}T_\mathrm{D}R_\mathrm{D}}}{\gamma_\mathrm{CCC}\sqrt{(1-L_\mathrm{D}C_\mathrm{D}(2\pi f)^2)^2+(R_\mathrm{D}C_\mathrm{D}2\pi f})^2}$ characterized by two main frequency ranges only, because $1/R_\mathrm{D}C_\mathrm{D}$ is close to $1/\sqrt{L_\mathrm{D}C_\mathrm{D}}$: i) for $f\ll 1/(2\pi R_\mathrm{D}C_\mathrm{D})$, $|S^\mathrm{D}_{\phi}(f)|=\frac{N_\mathrm{D}C_\mathrm{D}2\pi f\sqrt{4k_\mathrm{B}T_\mathrm{D}R_\mathrm{D}}}{\gamma_\mathrm{CCC}}$, ii) for $f \gg 1/(2\pi\sqrt{L_\mathrm{D}C_\mathrm{D}})$, $|S^\mathrm{D}_{\phi}(f)|=\frac{N_\mathrm{D}\sqrt{4k_\mathrm{B}T_\mathrm{D}R_\mathrm{D}}}{\gamma_\mathrm{CCC}L_\mathrm{D}2\pi f}$. Fig.2d shows that the $|S^\mathrm{D}_{\phi}(f)|$ fitting function (blue dashed line) adjusts very well the experimental detected noise (red) in the frequency range from 10 Hz up to 10 kHz. This extra noise manifests itself differently in the measurements according to the frequency bandwidth of the detector.
\subsection{\label{sec:level1}Type B standard uncertainty budget of the PQCG}
Table II presents the different contributions to the Type B uncertainty of the PQCG current that were evaluated: the cable correction $\alpha$ (Appendix D), the feedback electronics (Appendix E), the CCC accuracy (Appendix C.3), the QHRS accuracy (Appendix C.2), current leakage (Section II), the frequency accuracy, and the calibration of the current divider fraction $\beta$ (Section II). The two main contributions comes from the cable correction and the current divider calibration. In principle, the PQCG should be implemented without using the current divider, as it was for the ammeter calibration, adjusting the current with the number of biased junctions $n_\mathrm{J}$ or the frequency only. In this case, the total Type B relative uncertainty, essentially caused by the cable correction, is evaluated to about $2.5\times10^{-9}$ . Let us note that this contribution should be cancelled by the implementation of a triple connection of the QHRS, as planned in the future. In the accuracy test, the current divider is used and gives a contribution to the Type B uncertainty from $2\times10^{-9}$ ($N_\mathrm{Div}/N=1$) to $8\times10^{-9}$ ($N_\mathrm{Div}/N=4$).
\begin{table}[htbp]
\begin{center}
\begin{tabular}{|l|c|c|c|c|}
  \hline
 \textbf{Contribution}&\textbf{$u$}&\textbf{Sensitivity}&\textbf{$u^\mathrm{B}_\mathrm{PQCG}$}\\
                      &\textbf{($10^{-9}$)}          &                    &\textbf{($10^{-9}$)}\\ \hline
  Cable correction &$u_\alpha=2.5$ &1&2.5 \\
  Electronic feedback &$<0.5\times10^{-9}$&1&$<0.5\times10^{-9}$ \\
  CCC accuracy &$<1\times10^{-10}$&1&$<1\times10^{-10}$ \\
  QHRS& $<1\times10^{-10}$&1&$<1\times10^{-10}$ \\
  Current leakage &$<1\times10^{-11}$&1&$<1\times10^{-11}$ \\
  Frequency &$<1\times10^{-11}$&1&$<1\times10^{-11}$ \\
  \textcolor{blue}{Current divider (CD)} & \textcolor{blue}{$u_{\beta}=0.5$} & \textcolor{blue}{$N_{Div}/N$}&\textcolor{blue}{$0.5\times N_{Div}/N$}\\ \hline
  Total (without CD)  & & &\textbf{2.5} \\ \hline
  Total \textcolor{blue}{(with CD)} & & &\textcolor{blue}{\textbf{8.4 ($N=1$)}}\\
  \hline
\end{tabular}
\caption{Type B standard uncertainty budget of PQCG \label{Type B PQCG}}
\end{center}
\end{table}
\subsection{\label{sec:level1}Determination of $\beta_0$ realizing equilibrium}
 For the accuracy experiments of PQCG, the current divider was used to adjust $I_\mathrm{PQCG}$ so that the voltage balance is realised. At equilibrium, $V_\mathrm{J}^{ref}/R_\mathrm{100}$ can then be compared to its theoretical value $\frac{N_\mathrm{JK}}{N+\beta_0 N_\mathrm{Div}}I_\mathrm{PQCS}$, where $\beta_0$ is the fraction of $I_\mathrm{PQCG}$ injected by the divider in the $N_\mathrm{Div}$-turns winding. In practice, to simplify the calibration measurement chain, two sets of non-zero voltages $\Delta V_1$ and $\Delta V_2$ obtained for two fractions $\beta_1$ and $\beta_2$, respectively below and above $\beta_0$ are measured (see Fig.3b). $\beta_0$ is given by $\beta_0=\beta_1+(\beta_2-\beta_1)\times\Delta V_1/(|\Delta V_1|+|\Delta V_2|)$. Depending of the measurement, $\beta_0$ was between a few $10^{-6}$ up to a few $10^{-4}$. Its uncertainty $u_{\beta_0}$ is given by $u^2_{\beta_0}= [(u_{\beta_1}^2|\Delta V_2|^2+u_{\beta_2}^2|\Delta V_1|^2)/(|\Delta V_1|+|\Delta V_2|)^2]+[(\beta_2-\beta_1)^2(|\Delta V_2|^2 u^2_{\Delta V_1}+|\Delta V_1|^2 u^2_{\Delta V_2})/(|\Delta V_1|+|\Delta V_2|)^4]$, where $u_{\Delta V_1}$ and $u_{\Delta V_2}$ are the experimental standard deviations of the mean (Type A evaluation) calculated from the measurements of $\Delta V_1$ and $\Delta V_2$, $u_{\beta_1}$ and $u_{\beta_2}$ are the calibration standard uncertainties of $\beta_1$ and $\beta_2$. Taking into account that $|\Delta V_1|\simeq |\Delta V_2|$ and that $\beta_1$ and $\beta_2$ are strongly correlated quantities, the first term contributes by $0.5\times10^{-9}$ to $u_{\beta_0}$.
\subsection{\label{sec:level1}Allan deviation}
The Allan variance\cite{Witt2005} is the 2-sample variance that relies on three hypotheses: the distribution law of data is normal, the power spectral density can be decomposed into powers of the frequency, the time between data is constant, equal to $\tau_0$, without dead time. The advantage of this variance over the classical variance is that it converges for most of the commonly encountered kinds of noise, whereas the classical variance does not always converge to a finite value.

Considering a measurement performed during a time $T=M\tau_0$, where $M$ is the total number of samples, and $\bar{q}_i$ the ith average of the samples calculated over an analysis time $\tau=m\tau_0$ where $m$ can be varied from up to $M/2$, the Allan variance is defined as:
\begin{center}$\sigma^2_q(\tau=m\tau_0)=\frac{1}{2(M-1)}\sum\limits_{i=1}^{M-1}(\bar{q}_{i+1}-\bar{q_{i}})^2$.\end{center}
The Allan variance can allow to differentiate noise types according to the exponent of its power dependence with time. As an example, white noise manifests itself by a $\tau^{-1}$ dependence. This corresponds to an Allan deviation, $\sigma_q(\tau=m\tau_0)$, characterized by  a $\tau^{-1/2}$ dependence. In this case, the Allan deviation is an unbiased estimator of the true deviation.

In the paper we have used the total Allan variance (TOTAVAR) calculated with the software STABLE 32 Version 1.5. The total Allan variance is similar to the Allan variance and has the same expected value, but offers improved confidence at long averaging times.
It is defined as:
\begin{center}$\sigma^2_q(\tau=m\tau_0)=\frac{1}{2(M-1)}\sum\limits_{i=1}^{M-1}(\bar{q}^{*}_{i+j+1}-\bar{q}^{*}_{i+j})^2$.\end{center}
where the $M$ samples measured at $\tau=\tau_0$ are extended by reflection at both ends to forma virtual array $\bar{q}^{*}$. The original data is in the center where $\bar{q}^{*}_{i}=\bar{q}_{i}$ for $i=1$ to $M$, and the extended data for $j=1$ to $M-1$ is equal to $\bar{q}^{*}_{1-j}=\bar{q}_{j}$ and $\bar{q}^{*}_{M+j}=\bar{q}_{M+1-j}$.
Let us remark that the total Allan variance can be calculated for an analysis time $\tau$ up to half the total measurement time.

Figure 3.c reports the relative total Allan deviation $\sigma(I_\mathrm{PQCG})/I_\mathrm{PQCG}$ or $\sigma(\Delta V)/V_\mathrm{J}^{ref}$, which was calculated from a series of 49 on-off-on cycles of $I_\mathrm{PQCG}$ using the software STABLE 32. The $\tau^{-1/2}$ dependence confirms the white noise of data and legitimates the calculation of the experimental deviation of the mean for the 11-cycles time series to evaluate the standard uncertainties $u_\mathrm{\Delta V}$ (Type A evaluation).
\subsection{\label{sec:level1}Measurement standard uncertainty of $\Delta I_\mathrm{PQCG}/I_\mathrm{PQCG}$ in accuracy quantization tests of the PQCG}
For the experiments consisting in testing the accuracy of the PQCG, the relative combined standard uncertainty $u(\Delta I_\mathrm{PQCG})/I_\mathrm{PQCG}$ of the relative deviation of the generated current to its theoretical expectation, $\Delta I_\mathrm{PQCG}/I_\mathrm{PQCG}$, is calculated from the combination, using the propagation law of uncertainties\cite{GUM}, from the cable correction $u_{\alpha}$ (Appendix D), the current divider fraction $u_{\beta_0}$ (Appendix H) realizing equilibrium (main contribution which combined Type A and Type B components) and the $100~\Omega$ resistor $uR_{100}/R_{100}=2.5\times10^{-9}$. It results that $u(\Delta I_\mathrm{PQCG})/I_\mathrm{PQCG}\simeq\sqrt{u_{\alpha}^2+(u_{\beta_0}\times N_\mathrm {Div}/N)^2+(uR_{100}/R_{100})^2}$. All uncertainties reported in figures of section I and III are combined standard uncertainties (1 s.d.).
\subsection{\label{sec:level1}Calibration of the DA: accuracy and measurement uncertainties}
The DA is a HP3458A multimeter. Prior to calibrations, the DA has been adjusted by using a 10 k$\Omega$ resistor standard and a 10 V Zener voltage standard calibrated in terms of $R_{\mathrm{K}}$ and $K_{\mathrm{J}}$ respectively. The manufacturer specifications of the apparatus concerning the accuracy of current measurements are the following: (10 ppm reading + 4 ppm range) in the 1 mA range, (10 ppm reading + 40 ppm range) in the $\mathrm{1~\mu A}$ range.
The 100 nF capacitance forms a low-pass filter with the input resistance of the DA resulting in cutoff frequencies $f_\mathrm{C}$ depending on the current range of the DA: $f_\mathrm{C}=16$ kHz for the 1 mA range($100~\Omega$ input resistance) and $f_\mathrm{C}=35$ Hz for $1~\mu A$ ($45.2~k\Omega$ input resistance).
The current values measured $I_\mathrm{Meas.}$ by the DA are compared to $I_\mathrm{PQCG}$ values that are calculated using $K_\mathrm{J-90}$ and $R_\mathrm{K-90}$ (Appendix A).
In Fig.4b, 4c and 4d, the errors bars correspond to relative Type A uncertainties evaluated by experimental standard deviations of the mean calculated from eight on-off-on cycles (about 15 minutes). In these experiments, the relative Type B uncertainty contribution of the PQCG, reduced to $u^\mathrm{B}_\mathrm{PQCG}=u_{\alpha}=2.5\times 10^{-9}$, is not included because it is negligible compared to the Type A uncertainty contribution. In Fig.4d, the relative uncertainties (red bars) of the DA are calculated from the dispersion of the uncertainties of several measurements performed with the DA entries connected to the 100 nF capacitance only and using the same protocol based on eight on-off-on cycles (about 15 minutes).



\begin{thebibliography}{65}%
\makeatletter
\providecommand \@ifxundefined [1]{%
 \@ifx{#1\undefined}
}%
\providecommand \@ifnum [1]{%
 \ifnum #1\expandafter \@firstoftwo
 \else \expandafter \@secondoftwo
 \fi
}%
\providecommand \@ifx [1]{%
 \ifx #1\expandafter \@firstoftwo
 \else \expandafter \@secondoftwo
 \fi
}%
\providecommand \natexlab [1]{#1}%
\providecommand \enquote  [1]{``#1''}%
\providecommand \bibnamefont  [1]{#1}%
\providecommand \bibfnamefont [1]{#1}%
\providecommand \citenamefont [1]{#1}%
\providecommand \href@noop [0]{\@secondoftwo}%
\providecommand \href [0]{\begingroup \@sanitize@url \@href}%
\providecommand \@href[1]{\@@startlink{#1}\@@href}%
\providecommand \@@href[1]{\endgroup#1\@@endlink}%
\providecommand \@sanitize@url [0]{\catcode `\\12\catcode `\$12\catcode
  `\&12\catcode `\#12\catcode `\^12\catcode `\_12\catcode `\%12\relax}%
\providecommand \@@startlink[1]{}%
\providecommand \@@endlink[0]{}%
\providecommand \url  [0]{\begingroup\@sanitize@url \@url }%
\providecommand \@url [1]{\endgroup\@href {#1}{\urlprefix }}%
\providecommand \urlprefix  [0]{URL }%
\providecommand \Eprint [0]{\href }%
\providecommand \doibase [0]{http://dx.doi.org/}%
\providecommand \selectlanguage [0]{\@gobble}%
\providecommand \bibinfo  [0]{\@secondoftwo}%
\providecommand \bibfield  [0]{\@secondoftwo}%
\providecommand \translation [1]{[#1]}%
\providecommand \BibitemOpen [0]{}%
\providecommand \bibitemStop [0]{}%
\providecommand \bibitemNoStop [0]{.\EOS\space}%
\providecommand \EOS [0]{\spacefactor3000\relax}%
\providecommand \BibitemShut  [1]{\csname bibitem#1\endcsname}%
\let\auto@bib@innerbib\@empty
\bibitem [{\citenamefont {BIPM}(8th edition, 2006)}]{BrochureSI}%
  \BibitemOpen
  \bibfield  {author} {\bibinfo {author} {\bibnamefont {BIPM}},\ }\href@noop {}
  {\emph {\bibinfo {title} {The International System of units (SI)}}}\
  (\bibinfo  {publisher} {http://www.bipm.org/en/si/, S\`evres},\ \bibinfo
  {year} {8th edition, 2006})\BibitemShut {NoStop}%
\bibitem [{\citenamefont {CIPM}(2014)}]{CGPM2014}%
  \BibitemOpen
  \bibfield  {author} {\bibinfo {author} {\bibnamefont {CIPM}},\ }\href@noop {}
  {\emph {\bibinfo {title} {Resolution 1 of the 25th CGPM}}}\ (\bibinfo {year}
  {2014})\BibitemShut {NoStop}%
\bibitem [{\citenamefont {Mills}\ \emph {et~al.}(2005)\citenamefont {Mills},
  \citenamefont {Mohr}, \citenamefont {Quinn}, \citenamefont {Taylor},\ and\
  \citenamefont {Williams}}]{Mills2005}%
  \BibitemOpen
  \bibfield  {author} {\bibinfo {author} {\bibfnamefont {I.~M.}\ \bibnamefont
  {Mills}}, \bibinfo {author} {\bibfnamefont {P.~J.}\ \bibnamefont {Mohr}},
  \bibinfo {author} {\bibfnamefont {T.~J.}\ \bibnamefont {Quinn}}, \bibinfo
  {author} {\bibfnamefont {B.~N.}\ \bibnamefont {Taylor}}, \ and\ \bibinfo
  {author} {\bibfnamefont {E.~R.}\ \bibnamefont {Williams}},\ }\bibfield
  {title} {\enquote {\bibinfo {title} {Redefinition of the kilogram: a decision
  whose time has come},}\ }\href@noop {} {\bibfield  {journal} {\bibinfo
  {journal} {Metrologia}\ }\textbf {\bibinfo {volume} {42}},\ \bibinfo {pages}
  {71--80} (\bibinfo {year} {2005})}\BibitemShut {NoStop}%
\bibitem [{\citenamefont {Mills}\ \emph {et~al.}(2006)\citenamefont {Mills},
  \citenamefont {Mohr}, \citenamefont {Quinn}, \citenamefont {Taylor},\ and\
  \citenamefont {Williams}}]{Mills2006}%
  \BibitemOpen
  \bibfield  {author} {\bibinfo {author} {\bibfnamefont {I.M.}\ \bibnamefont
  {Mills}}, \bibinfo {author} {\bibfnamefont {P.J.}\ \bibnamefont {Mohr}},
  \bibinfo {author} {\bibfnamefont {T.J.}\ \bibnamefont {Quinn}}, \bibinfo
  {author} {\bibfnamefont {B.N.}\ \bibnamefont {Taylor}}, \ and\ \bibinfo
  {author} {\bibfnamefont {E.R.}\ \bibnamefont {Williams}},\ }\bibfield
  {title} {\enquote {\bibinfo {title} {Redefinition of the kilogram, ampere,
  kelvin and mole: a proposed approach to implementing \textsc{CIPM}
  recommendation 1},}\ }\href@noop {} {\bibfield  {journal} {\bibinfo
  {journal} {Metrologia}\ }\textbf {\bibinfo {volume} {43}},\ \bibinfo {pages}
  {227--246} (\bibinfo {year} {2006})}\BibitemShut {NoStop}%
\bibitem [{\citenamefont {Milton}\ \emph {et~al.}(2007)\citenamefont {Milton},
  \citenamefont {Wiliams},\ and\ \citenamefont {Bennett}}]{Milton2007}%
  \BibitemOpen
  \bibfield  {author} {\bibinfo {author} {\bibfnamefont {M.~J.T.}\ \bibnamefont
  {Milton}}, \bibinfo {author} {\bibfnamefont {J.~M.}\ \bibnamefont {Wiliams}},
  \ and\ \bibinfo {author} {\bibfnamefont {S.~J.}\ \bibnamefont {Bennett}},\
  }\bibfield  {title} {\enquote {\bibinfo {title} {Modernizing the \textsc{SI}:
  towards an improved, accessible and enduring system},}\ }\href@noop {}
  {\bibfield  {journal} {\bibinfo  {journal} {Metrologia}\ }\textbf {\bibinfo
  {volume} {44}},\ \bibinfo {pages} {356--364} (\bibinfo {year}
  {2007})}\BibitemShut {NoStop}%
\bibitem [{\citenamefont {BIPM}(2015)}]{Draft2015}%
  \BibitemOpen
  \bibfield  {author} {\bibinfo {author} {\bibnamefont {BIPM}},\ }\href@noop {}
  {\emph {\bibinfo {title} {Draft of the ninth SI Brochure}}}\ (\bibinfo {year}
  {2015})\ p.~\bibinfo {pages} {4}\BibitemShut {NoStop}%
\bibitem [{\citenamefont {Kibble}(1976)}]{Kibble1976}%
  \BibitemOpen
  \bibfield  {author} {\bibinfo {author} {\bibfnamefont {B.P.}\ \bibnamefont
  {Kibble}},\ }\enquote {\bibinfo {title} {Atomic masses and fundamental
  constants, vol.5},}\ \ (\bibinfo  {publisher} {Plenum Press},\ \bibinfo
  {year} {1976})\ Chap.\ \bibinfo {chapter} {A measurement of the gyromagnetic
  ratio of the proton by the strong field method}, pp.\ \bibinfo {pages}
  {545--551}\BibitemShut {NoStop}%
\bibitem [{\citenamefont {Stock}(2013)}]{Stock2013}%
  \BibitemOpen
  \bibfield  {author} {\bibinfo {author} {\bibfnamefont {M.}~\bibnamefont
  {Stock}},\ }\bibfield  {title} {\enquote {\bibinfo {title} {Watt balance
  experiments for the determination of the planck constant and the redefinition
  of the kilogram},}\ }\href@noop {} {\bibfield  {journal} {\bibinfo  {journal}
  {Metrologia}\ }\textbf {\bibinfo {volume} {50}},\ \bibinfo {pages} {R1}
  (\bibinfo {year} {2013})}\BibitemShut {NoStop}%
\bibitem [{Amp()}]{AmpereDefinition}%
  \BibitemOpen
  \bibfield  {title} {\enquote {\bibinfo {title} {The ampere is that constant
  current which, if maintained in two straight parallel conductors of infinite
  length, of negligible circular cross-section, and placed 1 metre apart in
  vacuum, would produce between these conductors a force equal to $2.10^{-7}$
  newton per metre of length.}}\ }in\ \href@noop {} {\emph {\bibinfo
  {booktitle} {The International System of Units (SI)}}},\ \bibinfo {editor}
  {edited by\ \bibinfo {editor} {\bibnamefont {BIPM}}}\ (\bibinfo  {publisher}
  {8th edition, 2006})\ p.\ \bibinfo {pages} {113}\BibitemShut {NoStop}%
\bibitem [{\citenamefont {BIPM}(2006)}]{SIANNEXE2}%
  \BibitemOpen
  \bibfield  {author} {\bibinfo {author} {\bibnamefont {BIPM}},\ }\href@noop {}
  {\emph {\bibinfo {title} {The International System of units (SI), Appendix
  2}}}\ (\bibinfo  {publisher} {http://www.bipm.org/en/si/, S\`evres},\
  \bibinfo {year} {2006})\BibitemShut {NoStop}%
\bibitem [{\citenamefont {Pekola}\ \emph {et~al.}(2013)\citenamefont {Pekola},
  \citenamefont {Saira}, \citenamefont {Maisi}, \citenamefont {Kemppinen},
  \citenamefont {M$\mathrm{\ddot{o}}$tt$\mathrm{\ddot{o}}$nen}, \citenamefont
  {Pashkin},\ and\ \citenamefont {Averin}}]{Pekola2013}%
  \BibitemOpen
  \bibfield  {author} {\bibinfo {author} {\bibfnamefont {J.~P.}\ \bibnamefont
  {Pekola}}, \bibinfo {author} {\bibfnamefont {O.P.}\ \bibnamefont {Saira}},
  \bibinfo {author} {\bibfnamefont {V.}~\bibnamefont {Maisi}}, \bibinfo
  {author} {\bibfnamefont {A.}~\bibnamefont {Kemppinen}}, \bibinfo {author}
  {\bibfnamefont {M.}~\bibnamefont
  {M$\mathrm{\ddot{o}}$tt$\mathrm{\ddot{o}}$nen}}, \bibinfo {author}
  {\bibfnamefont {Y.~A.}\ \bibnamefont {Pashkin}}, \ and\ \bibinfo {author}
  {\bibfnamefont {D.}~\bibnamefont {Averin}},\ }\bibfield  {title} {\enquote
  {\bibinfo {title} {Single-electron current sources: Toward a refined
  definition of the ampere},}\ }\href@noop {} {\bibfield  {journal} {\bibinfo
  {journal} {Rev. Mod. Phys.}\ }\textbf {\bibinfo {volume} {85}},\ \bibinfo
  {pages} {1421--1472} (\bibinfo {year} {2013})}\BibitemShut {NoStop}%
\bibitem [{\citenamefont {Grabert}\ and\ \citenamefont
  {Devoret}(1991)}]{Grabert1991}%
  \BibitemOpen
  \bibfield  {author} {\bibinfo {author} {\bibfnamefont {H.}~\bibnamefont
  {Grabert}}\ and\ \bibinfo {author} {\bibfnamefont {M.~H.}\ \bibnamefont
  {Devoret}},\ }\bibfield  {title} {\enquote {\bibinfo {title} {Single charge
  tunneling coulomb blockade phenomena in nanostructures},}\ \ }(\bibinfo
  {publisher} {Plenum Press},\ \bibinfo {year} {1991})\BibitemShut {NoStop}%
\bibitem [{\citenamefont {Pothier}\ \emph {et~al.}(1992)\citenamefont
  {Pothier}, \citenamefont {Lafarge}, \citenamefont {Urbina}, \citenamefont
  {Est\`eve},\ and\ \citenamefont {Devoret}}]{Pothier1992}%
  \BibitemOpen
  \bibfield  {author} {\bibinfo {author} {\bibfnamefont {H.}~\bibnamefont
  {Pothier}}, \bibinfo {author} {\bibfnamefont {P.}~\bibnamefont {Lafarge}},
  \bibinfo {author} {\bibfnamefont {C.}~\bibnamefont {Urbina}}, \bibinfo
  {author} {\bibfnamefont {D.}~\bibnamefont {Est\`eve}}, \ and\ \bibinfo
  {author} {\bibfnamefont {M.~H.}\ \bibnamefont {Devoret}},\ }\bibfield
  {title} {\enquote {\bibinfo {title} {Single-electron pump based on charging
  effects},}\ }\href@noop {} {\bibfield  {journal} {\bibinfo  {journal} {Eur.
  Phys. Lett.}\ }\textbf {\bibinfo {volume} {17}},\ \bibinfo {pages} {249}
  (\bibinfo {year} {1992})}\BibitemShut {NoStop}%
\bibitem [{\citenamefont {Keller}\ \emph {et~al.}(1996)\citenamefont {Keller},
  \citenamefont {Martinis}, \citenamefont {Zimmerman},\ and\ \citenamefont
  {Steinbach}}]{Keller1996}%
  \BibitemOpen
  \bibfield  {author} {\bibinfo {author} {\bibfnamefont {M.W.}\ \bibnamefont
  {Keller}}, \bibinfo {author} {\bibfnamefont {J.M.}\ \bibnamefont {Martinis}},
  \bibinfo {author} {\bibfnamefont {N.M.}\ \bibnamefont {Zimmerman}}, \ and\
  \bibinfo {author} {\bibfnamefont {A.H.}\ \bibnamefont {Steinbach}},\
  }\bibfield  {title} {\enquote {\bibinfo {title} {Accuracy of electron
  counting using a 7-junction electron pump},}\ }\href@noop {} {\bibfield
  {journal} {\bibinfo  {journal} {Appl. Phys. Lett.}\ }\textbf {\bibinfo
  {volume} {69}},\ \bibinfo {pages} {1804--1806} (\bibinfo {year}
  {1996})}\BibitemShut {NoStop}%
\bibitem [{\citenamefont {Keller}\ \emph {et~al.}(1999)\citenamefont {Keller},
  \citenamefont {Eichenberger}, \citenamefont {Martinis},\ and\ \citenamefont
  {Zimmermann}}]{Keller1999}%
  \BibitemOpen
  \bibfield  {author} {\bibinfo {author} {\bibfnamefont {M.W.}\ \bibnamefont
  {Keller}}, \bibinfo {author} {\bibfnamefont {A.L.}\ \bibnamefont
  {Eichenberger}}, \bibinfo {author} {\bibfnamefont {J.M.}\ \bibnamefont
  {Martinis}}, \ and\ \bibinfo {author} {\bibfnamefont {N.M.}\ \bibnamefont
  {Zimmermann}},\ }\bibfield  {title} {\enquote {\bibinfo {title} {A
  capacitance standard based on counting electrons},}\ }\href@noop {}
  {\bibfield  {journal} {\bibinfo  {journal} {Science}\ }\textbf {\bibinfo
  {volume} {285}},\ \bibinfo {pages} {1706} (\bibinfo {year}
  {1999})}\BibitemShut {NoStop}%
\bibitem [{\citenamefont {Keller}\ \emph {et~al.}(2007)\citenamefont {Keller},
  \citenamefont {Zimmermann},\ and\ \citenamefont {Eichenberger}}]{Keller2007}%
  \BibitemOpen
  \bibfield  {author} {\bibinfo {author} {\bibfnamefont {M.W.}\ \bibnamefont
  {Keller}}, \bibinfo {author} {\bibfnamefont {N.~M.}\ \bibnamefont
  {Zimmermann}}, \ and\ \bibinfo {author} {\bibfnamefont {A.~L.}\ \bibnamefont
  {Eichenberger}},\ }\bibfield  {title} {\enquote {\bibinfo {title}
  {Uncertainty budget for the nist electron counting capacitance standard,
  eccs-1},}\ }\href@noop {} {\bibfield  {journal} {\bibinfo  {journal}
  {Metrologia}\ }\textbf {\bibinfo {volume} {44}},\ \bibinfo {pages} {505}
  (\bibinfo {year} {2007})}\BibitemShut {NoStop}%
\bibitem [{\citenamefont {Camarota}\ \emph {et~al.}(2012)\citenamefont
  {Camarota}, \citenamefont {Scherer}, \citenamefont {Keller}, \citenamefont
  {Lotkov}, \citenamefont {Willenberg},\ and\ \citenamefont
  {Ahlers}}]{Camarota2012}%
  \BibitemOpen
  \bibfield  {author} {\bibinfo {author} {\bibfnamefont {B.}~\bibnamefont
  {Camarota}}, \bibinfo {author} {\bibfnamefont {H.}~\bibnamefont {Scherer}},
  \bibinfo {author} {\bibfnamefont {M.W.}\ \bibnamefont {Keller}}, \bibinfo
  {author} {\bibfnamefont {S.V.}\ \bibnamefont {Lotkov}}, \bibinfo {author}
  {\bibfnamefont {G.}~\bibnamefont {Willenberg}}, \ and\ \bibinfo {author}
  {\bibfnamefont {J.}~\bibnamefont {Ahlers}},\ }\bibfield  {title} {\enquote
  {\bibinfo {title} {Electron counting capacitance standard with an improved
  five-junction \textsc{R}-pump},}\ }\href@noop {} {\bibfield  {journal}
  {\bibinfo  {journal} {Metrologia}\ }\textbf {\bibinfo {volume} {49}},\
  \bibinfo {pages} {8--14} (\bibinfo {year} {2012})}\BibitemShut {NoStop}%
\bibitem [{\citenamefont {Kaestner}\ and\ \citenamefont
  {Kashcheyevs}(2015)}]{Kaestner2015}%
  \BibitemOpen
  \bibfield  {author} {\bibinfo {author} {\bibfnamefont {B.}~\bibnamefont
  {Kaestner}}\ and\ \bibinfo {author} {\bibfnamefont {V.}~\bibnamefont
  {Kashcheyevs}},\ }\bibfield  {title} {\enquote {\bibinfo {title}
  {Non-adiabatic quantized charge pumping with tunable-barrier quantum dots: a
  review of current progress},}\ }\href@noop {} {\bibfield  {journal} {\bibinfo
   {journal} {Rep. Prog. Phys.}\ }\textbf {\bibinfo {volume} {78}},\ \bibinfo
  {pages} {103901} (\bibinfo {year} {2015})}\BibitemShut {NoStop}%
\bibitem [{\citenamefont {Giblin}\ \emph {et~al.}(2012)\citenamefont {Giblin},
  \citenamefont {Kataoka}, \citenamefont {Fletcher}, \citenamefont {See},
  \citenamefont {Janssen}, \citenamefont {Griffiths}, \citenamefont {Jones},
  \citenamefont {Farrer},\ and\ \citenamefont {Ritchie}}]{Giblin2012}%
  \BibitemOpen
  \bibfield  {author} {\bibinfo {author} {\bibfnamefont {S.~P.}\ \bibnamefont
  {Giblin}}, \bibinfo {author} {\bibfnamefont {M.}~\bibnamefont {Kataoka}},
  \bibinfo {author} {\bibfnamefont {J.~D.}\ \bibnamefont {Fletcher}}, \bibinfo
  {author} {\bibfnamefont {P.}~\bibnamefont {See}}, \bibinfo {author}
  {\bibfnamefont {T.~J. B.~M.}\ \bibnamefont {Janssen}}, \bibinfo {author}
  {\bibfnamefont {J.~P.}\ \bibnamefont {Griffiths}}, \bibinfo {author}
  {\bibfnamefont {G.~A.~C.}\ \bibnamefont {Jones}}, \bibinfo {author}
  {\bibfnamefont {I.}~\bibnamefont {Farrer}}, \ and\ \bibinfo {author}
  {\bibfnamefont {D.~A.}\ \bibnamefont {Ritchie}},\ }\bibfield  {title}
  {\enquote {\bibinfo {title} {Towards a quantum representation of the ampere
  using single electron pumps},}\ }\href@noop {} {\bibfield  {journal}
  {\bibinfo  {journal} {Nat. Commun.}\ }\textbf {\bibinfo {volume} {3}},\
  \bibinfo {pages} {930} (\bibinfo {year} {2012})}\BibitemShut {NoStop}%
\bibitem [{\citenamefont {Stein}\ \emph {et~al.}(2015)\citenamefont {Stein},
  \citenamefont {Drung}, \citenamefont {Fricke}, \citenamefont {Scherer},
  \citenamefont {Hohls}, \citenamefont {Leicht}, \citenamefont {G\"{o}tz},
  \citenamefont {Krause}, \citenamefont {Behr}, \citenamefont {Pesel},
  \citenamefont {Pierz}, \citenamefont {Siegner}, \citenamefont {Ahlers},\ and\
  \citenamefont {Schumacher}}]{Stein2015}%
  \BibitemOpen
  \bibfield  {author} {\bibinfo {author} {\bibfnamefont {F.}~\bibnamefont
  {Stein}}, \bibinfo {author} {\bibfnamefont {D.}~\bibnamefont {Drung}},
  \bibinfo {author} {\bibfnamefont {L.}~\bibnamefont {Fricke}}, \bibinfo
  {author} {\bibfnamefont {H.}~\bibnamefont {Scherer}}, \bibinfo {author}
  {\bibfnamefont {F.}~\bibnamefont {Hohls}}, \bibinfo {author} {\bibfnamefont
  {C.}~\bibnamefont {Leicht}}, \bibinfo {author} {\bibfnamefont
  {M.}~\bibnamefont {G\"{o}tz}}, \bibinfo {author} {\bibfnamefont
  {C.}~\bibnamefont {Krause}}, \bibinfo {author} {\bibfnamefont
  {R.}~\bibnamefont {Behr}}, \bibinfo {author} {\bibfnamefont {E.}~\bibnamefont
  {Pesel}}, \bibinfo {author} {\bibfnamefont {K.}~\bibnamefont {Pierz}},
  \bibinfo {author} {\bibfnamefont {U.}~\bibnamefont {Siegner}}, \bibinfo
  {author} {\bibfnamefont {F.~J.}\ \bibnamefont {Ahlers}}, \ and\ \bibinfo
  {author} {\bibfnamefont {H.~W.}\ \bibnamefont {Schumacher}},\ }\bibfield
  {title} {\enquote {\bibinfo {title} {Validation of a quantized-current source
  with 0.2 ppm uncertainty},}\ }\href@noop {} {\bibfield  {journal} {\bibinfo
  {journal} {Appl. Phys. Lett.}\ }\textbf {\bibinfo {volume} {107}},\ \bibinfo
  {pages} {103501} (\bibinfo {year} {2015})}\BibitemShut {NoStop}%
\bibitem [{\citenamefont {BIPM}(2016)}]{BestCMC}%
  \BibitemOpen
  \bibfield  {author} {\bibinfo {author} {\bibnamefont {BIPM}},\ }\href@noop {}
  {\emph {\bibinfo {title} {Calibration and Measurement Capabilities
  Electricity and Magnetism: DC Current}}}\ (\bibinfo  {publisher} {Key
  comparison database},\ \bibinfo {year} {2016})\BibitemShut {NoStop}%
\bibitem [{\citenamefont {Josephson}(1962)}]{Josephson62}%
  \BibitemOpen
  \bibfield  {author} {\bibinfo {author} {\bibfnamefont {B.D.}\ \bibnamefont
  {Josephson}},\ }\bibfield  {title} {\enquote {\bibinfo {title} {Possible new
  effects in superconductive tunnelling},}\ }\href@noop {} {\bibfield
  {journal} {\bibinfo  {journal} {Phys. Lett.}\ }\textbf {\bibinfo {volume}
  {1}},\ \bibinfo {pages} {251} (\bibinfo {year} {1962})}\BibitemShut {NoStop}%
\bibitem [{\citenamefont {Klitzing}\ \emph {et~al.}(1980)\citenamefont
  {Klitzing}, \citenamefont {Dorda},\ and\ \citenamefont
  {Pepper}}]{Klitzing1980}%
  \BibitemOpen
  \bibfield  {author} {\bibinfo {author} {\bibfnamefont {K.V.}\ \bibnamefont
  {Klitzing}}, \bibinfo {author} {\bibfnamefont {G.}~\bibnamefont {Dorda}}, \
  and\ \bibinfo {author} {\bibfnamefont {M.}~\bibnamefont {Pepper}},\
  }\bibfield  {title} {\enquote {\bibinfo {title} {New method for high-accuracy
  determination of the fine-structure constant based on quantized hall
  resistance},}\ }\href@noop {} {\bibfield  {journal} {\bibinfo  {journal}
  {Phys. Rev. Lett.}\ }\textbf {\bibinfo {volume} {45}},\ \bibinfo {pages}
  {494} (\bibinfo {year} {1980})}\BibitemShut {NoStop}%
\bibitem [{\citenamefont {Shapiro}(1963)}]{Shapiro63}%
  \BibitemOpen
  \bibfield  {author} {\bibinfo {author} {\bibfnamefont {S.}~\bibnamefont
  {Shapiro}},\ }\bibfield  {title} {\enquote {\bibinfo {title} {Josephson
  currents in superconducting tunneling: the effect of microwaves and another
  observations},}\ }\href@noop {} {\bibfield  {journal} {\bibinfo  {journal}
  {Phys. Rev. Lett.}\ }\textbf {\bibinfo {volume} {11}},\ \bibinfo {pages}
  {80--82} (\bibinfo {year} {1963})}\BibitemShut {NoStop}%
\bibitem [{\citenamefont {Buttiker}(1988)}]{Buttiker1988}%
  \BibitemOpen
  \bibfield  {author} {\bibinfo {author} {\bibfnamefont {M.}~\bibnamefont
  {Buttiker}},\ }\bibfield  {title} {\enquote {\bibinfo {title} {Absence of
  backscattering in the quantum hall effect in multiprobe conductors},}\
  }\href@noop {} {\bibfield  {journal} {\bibinfo  {journal} {Phys. Rev. B}\
  }\textbf {\bibinfo {volume} {38}},\ \bibinfo {pages} {9375} (\bibinfo {year}
  {1988})}\BibitemShut {NoStop}%
\bibitem [{\citenamefont {Bloch}(1970)}]{Bloch1970}%
  \BibitemOpen
  \bibfield  {author} {\bibinfo {author} {\bibfnamefont {F.}~\bibnamefont
  {Bloch}},\ }\bibfield  {title} {\enquote {\bibinfo {title} {Josephson effect
  in a superconducting ring},}\ }\href@noop {} {\bibfield  {journal} {\bibinfo
  {journal} {Phys. Rev. B}\ }\textbf {\bibinfo {volume} {2}},\ \bibinfo {pages}
  {109} (\bibinfo {year} {1970})}\BibitemShut {NoStop}%
\bibitem [{\citenamefont {Laughlin}(1981)}]{Laughlin81}%
  \BibitemOpen
  \bibfield  {author} {\bibinfo {author} {\bibfnamefont {R.~B.}\ \bibnamefont
  {Laughlin}},\ }\bibfield  {title} {\enquote {\bibinfo {title} {Quantized
  \textsc{H}all conductivity in two dimensions},}\ }\href@noop {} {\bibfield
  {journal} {\bibinfo  {journal} {Phys. Rev. B}\ }\textbf {\bibinfo {volume}
  {23}},\ \bibinfo {pages} {5632--5633} (\bibinfo {year} {1981})}\BibitemShut
  {NoStop}%
\bibitem [{\citenamefont {Thouless}(1994)}]{Thouless1994}%
  \BibitemOpen
  \bibfield  {author} {\bibinfo {author} {\bibfnamefont {D.~J.}\ \bibnamefont
  {Thouless}},\ }\bibfield  {title} {\enquote {\bibinfo {title} {Topological
  interpretations of quantum hall conductance},}\ }\href@noop {} {\bibfield
  {journal} {\bibinfo  {journal} {J. Math. Phys.}\ }\textbf {\bibinfo {volume}
  {35}},\ \bibinfo {pages} {5362} (\bibinfo {year} {1994})}\BibitemShut
  {NoStop}%
\bibitem [{\citenamefont {Penin}(2009)}]{Penin2009}%
  \BibitemOpen
  \bibfield  {author} {\bibinfo {author} {\bibfnamefont {A.~A.}\ \bibnamefont
  {Penin}},\ }\bibfield  {title} {\enquote {\bibinfo {title} {Quantum hall
  effect in quantum electrodynamics},}\ }\href@noop {} {\bibfield  {journal}
  {\bibinfo  {journal} {Phys. Rev. B.}\ }\textbf {\bibinfo {volume} {79}},\
  \bibinfo {pages} {113303} (\bibinfo {year} {2009})},\ \bibinfo {note}
  {erratum Phys. Rev. B 81, 089902 (2010)}\BibitemShut {NoStop}%
\bibitem [{\citenamefont {Penin}(2010)}]{Penin2010}%
  \BibitemOpen
  \bibfield  {author} {\bibinfo {author} {\bibfnamefont {A.~A.}\ \bibnamefont
  {Penin}},\ }\bibfield  {title} {\enquote {\bibinfo {title} {Measuring vacuum
  polarization with josephson junctions},}\ }\href@noop {} {\bibfield
  {journal} {\bibinfo  {journal} {Phys. Rev. Lett.}\ }\textbf {\bibinfo
  {volume} {104}},\ \bibinfo {pages} {097003} (\bibinfo {year}
  {2010})}\BibitemShut {NoStop}%
\bibitem [{\citenamefont {Mohr}\ \emph {et~al.}(2012)\citenamefont {Mohr},
  \citenamefont {Taylor},\ and\ \citenamefont {Newell}}]{MohrCODATA2010}%
  \BibitemOpen
  \bibfield  {author} {\bibinfo {author} {\bibfnamefont {P.~J.}\ \bibnamefont
  {Mohr}}, \bibinfo {author} {\bibfnamefont {B.~N.}\ \bibnamefont {Taylor}}, \
  and\ \bibinfo {author} {\bibfnamefont {D.~B.}\ \bibnamefont {Newell}},\
  }\bibfield  {title} {\enquote {\bibinfo {title} {\textsc{CODATA} recommended
  values of the fundamental physical constants: 2010},}\ }\href@noop {}
  {\bibfield  {journal} {\bibinfo  {journal} {Rev. Mod. Phys.}\ }\textbf
  {\bibinfo {volume} {84}},\ \bibinfo {pages} {1527--1605} (\bibinfo {year}
  {2012})}\BibitemShut {NoStop}%
\bibitem [{\citenamefont {Mohr}\ \emph {et~al.}(2016)\citenamefont {Mohr},
  \citenamefont {Newell},\ and\ \citenamefont {Taylor}}]{codata16}%
  \BibitemOpen
  \bibfield  {author} {\bibinfo {author} {\bibfnamefont {P.J.}\ \bibnamefont
  {Mohr}}, \bibinfo {author} {\bibfnamefont {D.B.}\ \bibnamefont {Newell}}, \
  and\ \bibinfo {author} {\bibfnamefont {B.N.}\ \bibnamefont {Taylor}},\
  }\bibfield  {title} {\enquote {\bibinfo {title} {Codata recommended values of
  the fundamental physical constants: 2014},}\ }\href@noop {} {\bibfield
  {journal} {\bibinfo  {journal} {Rev. Mod. Phys.}\ }\textbf {\bibinfo {volume}
  {88}} (\bibinfo {year} {2016})}\BibitemShut {NoStop}%
\bibitem [{\citenamefont {Tsai}\ \emph {et~al.}(1983)\citenamefont {Tsai},
  \citenamefont {Jain},\ and\ \citenamefont {Lukens}}]{Tsai1983}%
  \BibitemOpen
  \bibfield  {author} {\bibinfo {author} {\bibfnamefont {J.~S.}\ \bibnamefont
  {Tsai}}, \bibinfo {author} {\bibfnamefont {A.~K.}\ \bibnamefont {Jain}}, \
  and\ \bibinfo {author} {\bibfnamefont {J.~E.}\ \bibnamefont {Lukens}},\
  }\bibfield  {title} {\enquote {\bibinfo {title} {High-precision test of the
  universality of the josephson voltage-frequency relation},}\ }\href@noop {}
  {\bibfield  {journal} {\bibinfo  {journal} {Phys. Rev. Lett.}\ }\textbf
  {\bibinfo {volume} {51}},\ \bibinfo {pages} {316} (\bibinfo {year}
  {1983})}\BibitemShut {NoStop}%
\bibitem [{\citenamefont {Jain}\ \emph {et~al.}(1987)\citenamefont {Jain},
  \citenamefont {Lukens},\ and\ \citenamefont {Tsai}}]{Jain1987}%
  \BibitemOpen
  \bibfield  {author} {\bibinfo {author} {\bibfnamefont {A.~K.}\ \bibnamefont
  {Jain}}, \bibinfo {author} {\bibfnamefont {J.~E.}\ \bibnamefont {Lukens}}, \
  and\ \bibinfo {author} {\bibfnamefont {J.~S}\ \bibnamefont {Tsai}},\
  }\bibfield  {title} {\enquote {\bibinfo {title} {Test for relativistic
  gravitational effects on charged particles},}\ }\href@noop {} {\bibfield
  {journal} {\bibinfo  {journal} {Phys. Rev. Lett.}\ }\textbf {\bibinfo
  {volume} {58}},\ \bibinfo {pages} {1165} (\bibinfo {year}
  {1987})}\BibitemShut {NoStop}%
\bibitem [{\citenamefont {Krasnopolin}\ \emph {et~al.}(2002)\citenamefont
  {Krasnopolin}, \citenamefont {Behr},\ and\ \citenamefont
  {Niemeyer}}]{Krasnopolin2002}%
  \BibitemOpen
  \bibfield  {author} {\bibinfo {author} {\bibfnamefont {I.~Y.}\ \bibnamefont
  {Krasnopolin}}, \bibinfo {author} {\bibfnamefont {R.}~\bibnamefont {Behr}}, \
  and\ \bibinfo {author} {\bibfnamefont {J.}~\bibnamefont {Niemeyer}},\
  }\bibfield  {title} {\enquote {\bibinfo {title} {Highly precise comparison of
  \textsc{N}b/\textsc{A}l/\textsc{A}lox/\textsc{A}l/\textsc{A}l\textsc{O}x/\textsc{A}l/\textsc{N}b
  josephson junction arrays using a \textsc{SQUID} as a null detector},}\
  }\href@noop {} {\bibfield  {journal} {\bibinfo  {journal} {Supercond. Sci.
  Technol.}\ }\textbf {\bibinfo {volume} {15}} (\bibinfo {year}
  {2002})}\BibitemShut {NoStop}%
\bibitem [{\citenamefont {Schopfer}\ and\ \citenamefont
  {Poirier}(2013)}]{Schopfer2013}%
  \BibitemOpen
  \bibfield  {author} {\bibinfo {author} {\bibfnamefont {F.}~\bibnamefont
  {Schopfer}}\ and\ \bibinfo {author} {\bibfnamefont {W.}~\bibnamefont
  {Poirier}},\ }\bibfield  {title} {\enquote {\bibinfo {title} {Quantum
  resistance standard accuracy close to the zero-dissipation state},}\
  }\href@noop {} {\bibfield  {journal} {\bibinfo  {journal} {J. Appl. Phys.}\
  }\textbf {\bibinfo {volume} {114}},\ \bibinfo {pages} {064508} (\bibinfo
  {year} {2013})}\BibitemShut {NoStop}%
\bibitem [{\citenamefont {Ribeiro-Palau}\ \emph {et~al.}(2015)\citenamefont
  {Ribeiro-Palau}, \citenamefont {Lafont}, \citenamefont {Brun-Picard},
  \citenamefont {Kazazis}, \citenamefont {Michon}, \citenamefont {Cheynis},
  \citenamefont {Couturaud}, \citenamefont {Consejo}, \citenamefont {Jouault},
  \citenamefont {Poirier},\ and\ \citenamefont {Schopfer}}]{Ribeiro2015}%
  \BibitemOpen
  \bibfield  {author} {\bibinfo {author} {\bibfnamefont {R.}~\bibnamefont
  {Ribeiro-Palau}}, \bibinfo {author} {\bibfnamefont {F.}~\bibnamefont
  {Lafont}}, \bibinfo {author} {\bibfnamefont {J.}~\bibnamefont {Brun-Picard}},
  \bibinfo {author} {\bibfnamefont {D.}~\bibnamefont {Kazazis}}, \bibinfo
  {author} {\bibfnamefont {A.}~\bibnamefont {Michon}}, \bibinfo {author}
  {\bibfnamefont {F.}~\bibnamefont {Cheynis}}, \bibinfo {author} {\bibfnamefont
  {O.}~\bibnamefont {Couturaud}}, \bibinfo {author} {\bibfnamefont
  {C.}~\bibnamefont {Consejo}}, \bibinfo {author} {\bibfnamefont
  {B.}~\bibnamefont {Jouault}}, \bibinfo {author} {\bibfnamefont
  {W.}~\bibnamefont {Poirier}}, \ and\ \bibinfo {author} {\bibfnamefont
  {F.}~\bibnamefont {Schopfer}},\ }\bibfield  {title} {\enquote {\bibinfo
  {title} {Quantum hall resistance standard in graphene devices under relaxed
  experimental conditions},}\ }\href@noop {} {\bibfield  {journal} {\bibinfo
  {journal} {Nature Nano.}\ }\textbf {\bibinfo {volume} {10}} (\bibinfo {year}
  {2015})}\BibitemShut {NoStop}%
\bibitem [{\citenamefont {Janssen}\ \emph {et~al.}(2011)\citenamefont
  {Janssen}, \citenamefont {Fletcher}, \citenamefont {Goebel}, \citenamefont
  {Williams}, \citenamefont {Tzalenchuk}, \citenamefont {Yakimova},
  \citenamefont {Kubatkin}, \citenamefont {Lara-Avila},\ and\ \citenamefont
  {Fal�ko}}]{Janssen2011}%
  \BibitemOpen
  \bibfield  {author} {\bibinfo {author} {\bibfnamefont {T.~J. B.~M.}\
  \bibnamefont {Janssen}}, \bibinfo {author} {\bibfnamefont {N.E.}\
  \bibnamefont {Fletcher}}, \bibinfo {author} {\bibfnamefont {R.}~\bibnamefont
  {Goebel}}, \bibinfo {author} {\bibfnamefont {J.M.}\ \bibnamefont {Williams}},
  \bibinfo {author} {\bibfnamefont {A.}~\bibnamefont {Tzalenchuk}}, \bibinfo
  {author} {\bibfnamefont {R.}~\bibnamefont {Yakimova}}, \bibinfo {author}
  {\bibfnamefont {S.}~\bibnamefont {Kubatkin}}, \bibinfo {author}
  {\bibfnamefont {S.}~\bibnamefont {Lara-Avila}}, \ and\ \bibinfo {author}
  {\bibfnamefont {V.I.}\ \bibnamefont {Fal�ko}},\ }\bibfield  {title}
  {\enquote {\bibinfo {title} {Graphene, universality of the quantum hall
  effect and redefinition of the si system},}\ }\href@noop {} {\bibfield
  {journal} {\bibinfo  {journal} {New J. Phys.}\ }\textbf {\bibinfo {volume}
  {13}},\ \bibinfo {pages} {093026} (\bibinfo {year} {2011})}\BibitemShut
  {NoStop}%
\bibitem [{\citenamefont {Quinn}(1989)}]{KJ}%
  \BibitemOpen
  \bibfield  {author} {\bibinfo {author} {\bibfnamefont {T.}~\bibnamefont
  {Quinn}},\ }\bibfield  {title} {\enquote {\bibinfo {title} {News from the
  bipm},}\ }\href@noop {} {\bibfield  {journal} {\bibinfo  {journal}
  {Metrologia}\ }\textbf {\bibinfo {volume} {26}},\ \bibinfo {pages} {69}
  (\bibinfo {year} {1989})}\BibitemShut {NoStop}%
\bibitem [{\citenamefont {CIPM}(2000)}]{RK}%
  \BibitemOpen
  \bibfield  {author} {\bibinfo {author} {\bibnamefont {CIPM}},\ }\bibfield
  {title} {\enquote {\bibinfo {title} {use of the von klitzing constant to
  express the value of a reference standard of resistance as a function of the
  quantum hall effect},}\ }\href@noop {} {\bibfield  {journal} {\bibinfo
  {journal} {Proc\`es-Verbaux des S\'{e}ances du Comit\'{e} International des
  Poids et Mesures}\ }\textbf {\bibinfo {volume} {89th meeting}},\ \bibinfo
  {pages} {101} (\bibinfo {year} {2000})}\BibitemShut {NoStop}%
\bibitem [{\citenamefont {Drung}\ \emph
  {et~al.}(2015{\natexlab{a}})\citenamefont {Drung}, \citenamefont {Krause},
  \citenamefont {Becker}, \citenamefont {Scherer},\ and\ \citenamefont
  {Ahlers}}]{DrungRSI2015}%
  \BibitemOpen
  \bibfield  {author} {\bibinfo {author} {\bibfnamefont {D.}~\bibnamefont
  {Drung}}, \bibinfo {author} {\bibfnamefont {C.}~\bibnamefont {Krause}},
  \bibinfo {author} {\bibfnamefont {U.}~\bibnamefont {Becker}}, \bibinfo
  {author} {\bibfnamefont {H.}~\bibnamefont {Scherer}}, \ and\ \bibinfo
  {author} {\bibfnamefont {F.J.}\ \bibnamefont {Ahlers}},\ }\bibfield  {title}
  {\enquote {\bibinfo {title} {Ultrasatable low-noise current amplifier: A
  novel device for measuring small electric currents with high accuracy},}\
  }\href@noop {} {\bibfield  {journal} {\bibinfo  {journal} {Rev. Sci.
  Instrum.}\ }\textbf {\bibinfo {volume} {86}},\ \bibinfo {pages} {024703}
  (\bibinfo {year} {2015}{\natexlab{a}})}\BibitemShut {NoStop}%
\bibitem [{\citenamefont {Drung}\ \emph
  {et~al.}(2015{\natexlab{b}})\citenamefont {Drung}, \citenamefont {Krause},
  \citenamefont {Giblin}, \citenamefont {Djordjevic}, \citenamefont {Piquemal},
  \citenamefont {S\'{e}ron}, \citenamefont {Rengnez}, \citenamefont {G\"{o}tz},
  \citenamefont {Pesel},\ and\ \citenamefont {Scherer}}]{Drung2015}%
  \BibitemOpen
  \bibfield  {author} {\bibinfo {author} {\bibfnamefont {D.}~\bibnamefont
  {Drung}}, \bibinfo {author} {\bibfnamefont {C.}~\bibnamefont {Krause}},
  \bibinfo {author} {\bibfnamefont {S.~P.}\ \bibnamefont {Giblin}}, \bibinfo
  {author} {\bibfnamefont {S.}~\bibnamefont {Djordjevic}}, \bibinfo {author}
  {\bibfnamefont {F.}~\bibnamefont {Piquemal}}, \bibinfo {author}
  {\bibfnamefont {O.}~\bibnamefont {S\'{e}ron}}, \bibinfo {author}
  {\bibfnamefont {F.}~\bibnamefont {Rengnez}}, \bibinfo {author} {\bibfnamefont
  {M.}~\bibnamefont {G\"{o}tz}}, \bibinfo {author} {\bibfnamefont
  {E.}~\bibnamefont {Pesel}}, \ and\ \bibinfo {author} {\bibfnamefont
  {H.}~\bibnamefont {Scherer}},\ }\bibfield  {title} {\enquote {\bibinfo
  {title} {Validation of the ultrastable low-noise current amplifier as
  travelling standard for small direct currents},}\ }\href@noop {} {\bibfield
  {journal} {\bibinfo  {journal} {Metrologia}\ }\textbf {\bibinfo {volume}
  {52}},\ \bibinfo {pages} {756} (\bibinfo {year}
  {2015}{\natexlab{b}})}\BibitemShut {NoStop}%
\bibitem [{\citenamefont {Likharev}\ and\ \citenamefont
  {Zorin}(1985)}]{Likharev1985}%
  \BibitemOpen
  \bibfield  {author} {\bibinfo {author} {\bibfnamefont {K.~K.}\ \bibnamefont
  {Likharev}}\ and\ \bibinfo {author} {\bibfnamefont {A.~B.}\ \bibnamefont
  {Zorin}},\ }\bibfield  {title} {\enquote {\bibinfo {title} {Theory of the
  bloch-wave oscillations in small josephson junctions},}\ }\href@noop {}
  {\bibfield  {journal} {\bibinfo  {journal} {J. Low Temp. Phys.}\ }\textbf
  {\bibinfo {volume} {59}},\ \bibinfo {pages} {347} (\bibinfo {year}
  {1985})}\BibitemShut {NoStop}%
\bibitem [{\citenamefont {Behr}\ \emph {et~al.}(2012)\citenamefont {Behr},
  \citenamefont {Kieler}, \citenamefont {Kohlmann}, \citenamefont
  {M\"{u}ller},\ and\ \citenamefont {Palafox}}]{BehrMST2012}%
  \BibitemOpen
  \bibfield  {author} {\bibinfo {author} {\bibfnamefont {R.}~\bibnamefont
  {Behr}}, \bibinfo {author} {\bibfnamefont {O.}~\bibnamefont {Kieler}},
  \bibinfo {author} {\bibfnamefont {J.}~\bibnamefont {Kohlmann}}, \bibinfo
  {author} {\bibfnamefont {F.}~\bibnamefont {M\"{u}ller}}, \ and\ \bibinfo
  {author} {\bibfnamefont {L.}~\bibnamefont {Palafox}},\ }\bibfield  {title}
  {\enquote {\bibinfo {title} {Development and metrological applications of
  josephson arrays at ptb},}\ }\href@noop {} {\bibfield  {journal} {\bibinfo
  {journal} {Meas. Sci. Tech.}\ }\textbf {\bibinfo {volume} {23}},\ \bibinfo
  {pages} {124002} (\bibinfo {year} {2012})}\BibitemShut {NoStop}%
\bibitem [{\citenamefont {Jeckelmann}\ and\ \citenamefont
  {Jeanneret}(2001)}]{Jeckelmann2001}%
  \BibitemOpen
  \bibfield  {author} {\bibinfo {author} {\bibfnamefont {B.}~\bibnamefont
  {Jeckelmann}}\ and\ \bibinfo {author} {\bibfnamefont {B.}~\bibnamefont
  {Jeanneret}},\ }\bibfield  {title} {\enquote {\bibinfo {title} {The quantum
  hall effect as an electrical resistance standard},}\ }\href@noop {}
  {\bibfield  {journal} {\bibinfo  {journal} {Rep. Prog. Phys.}\ }\textbf
  {\bibinfo {volume} {64}},\ \bibinfo {pages} {1603} (\bibinfo {year}
  {2001})}\BibitemShut {NoStop}%
\bibitem [{\citenamefont {Poirier}\ and\ \citenamefont
  {Schopfer}(2009)}]{Poirier2009}%
  \BibitemOpen
  \bibfield  {author} {\bibinfo {author} {\bibfnamefont {W.}~\bibnamefont
  {Poirier}}\ and\ \bibinfo {author} {\bibfnamefont {F.}~\bibnamefont
  {Schopfer}},\ }\bibfield  {title} {\enquote {\bibinfo {title} {Resistance
  metrology based on the quantum hall effect},}\ }\href@noop {} {\bibfield
  {journal} {\bibinfo  {journal} {Eur. Phys. J. Spec. Top.}\ }\textbf {\bibinfo
  {volume} {172}},\ \bibinfo {pages} {207} (\bibinfo {year}
  {2009})}\BibitemShut {NoStop}%
\bibitem [{\citenamefont {Mueller}\ \emph {et~al.}(2007)\citenamefont
  {Mueller}, \citenamefont {Behr}, \citenamefont {Palafox}, \citenamefont
  {Kohlmann}, \citenamefont {Wendisch},\ and\ \citenamefont
  {Krasnopolin}}]{Mueller2007}%
  \BibitemOpen
  \bibfield  {author} {\bibinfo {author} {\bibfnamefont {F.}~\bibnamefont
  {Mueller}}, \bibinfo {author} {\bibfnamefont {R.}~\bibnamefont {Behr}},
  \bibinfo {author} {\bibfnamefont {L.}~\bibnamefont {Palafox}}, \bibinfo
  {author} {\bibfnamefont {J.}~\bibnamefont {Kohlmann}}, \bibinfo {author}
  {\bibfnamefont {R.}~\bibnamefont {Wendisch}}, \ and\ \bibinfo {author}
  {\bibfnamefont {I.}~\bibnamefont {Krasnopolin}},\ }\bibfield  {title}
  {\enquote {\bibinfo {title} {Improved 10 v sinis series arrays for
  applications in ac voltage metrology},}\ }\href@noop {} {\bibfield  {journal}
  {\bibinfo  {journal} {IEEE Trans. Appl. Supercond.}\ }\textbf {\bibinfo
  {volume} {17}},\ \bibinfo {pages} {649} (\bibinfo {year} {2007})}\BibitemShut
  {NoStop}%
\bibitem [{\citenamefont {Behr}\ \emph {et~al.}(2003)\citenamefont {Behr},
  \citenamefont {Kohlmann}, \citenamefont {Jansen}, \citenamefont
  {Kleinschmidt}, \citenamefont {Williams}, \citenamefont {Djordjevic},
  \citenamefont {Lo-Hive}, \citenamefont {Piquemal}, \citenamefont {Hetland},
  \citenamefont {Reymann}, \citenamefont {Eklund}, \citenamefont {Hof},
  \citenamefont {Jeanneret}, \citenamefont {Chevtchenko}, \citenamefont
  {Houtzager}, \citenamefont {van~den Brom}, \citenamefont {Sosso},
  \citenamefont {Andreone}, \citenamefont {Nassila},\ and\ \citenamefont
  {Helisto}}]{Behr2003}%
  \BibitemOpen
  \bibfield  {author} {\bibinfo {author} {\bibfnamefont {Ralf}\ \bibnamefont
  {Behr}}, \bibinfo {author} {\bibfnamefont {J.}~\bibnamefont {Kohlmann}},
  \bibinfo {author} {\bibfnamefont {J-T.}\ \bibnamefont {Jansen}}, \bibinfo
  {author} {\bibfnamefont {P.}~\bibnamefont {Kleinschmidt}}, \bibinfo {author}
  {\bibfnamefont {J.}~\bibnamefont {Williams}}, \bibinfo {author}
  {\bibfnamefont {S.}~\bibnamefont {Djordjevic}}, \bibinfo {author}
  {\bibfnamefont {J-P.}\ \bibnamefont {Lo-Hive}}, \bibinfo {author}
  {\bibfnamefont {F.}~\bibnamefont {Piquemal}}, \bibinfo {author}
  {\bibfnamefont {P.}~\bibnamefont {Hetland}}, \bibinfo {author} {\bibfnamefont
  {D.}~\bibnamefont {Reymann}}, \bibinfo {author} {\bibfnamefont
  {G.}~\bibnamefont {Eklund}}, \bibinfo {author} {\bibfnamefont
  {C.}~\bibnamefont {Hof}}, \bibinfo {author} {\bibfnamefont {B.}~\bibnamefont
  {Jeanneret}}, \bibinfo {author} {\bibfnamefont {O.}~\bibnamefont
  {Chevtchenko}}, \bibinfo {author} {\bibfnamefont {E.}~\bibnamefont
  {Houtzager}}, \bibinfo {author} {\bibfnamefont {H.}~\bibnamefont {van~den
  Brom}}, \bibinfo {author} {\bibfnamefont {A.}~\bibnamefont {Sosso}}, \bibinfo
  {author} {\bibfnamefont {D.}~\bibnamefont {Andreone}}, \bibinfo {author}
  {\bibfnamefont {J.}~\bibnamefont {Nassila}}, \ and\ \bibinfo {author}
  {\bibfnamefont {P.}~\bibnamefont {Helisto}},\ }\bibfield  {title} {\enquote
  {\bibinfo {title} {Analysis of different measurement setups for a
  programmable josephson voltage standard},}\ }\href@noop {} {\bibfield
  {journal} {\bibinfo  {journal} {IEEE Trans. Instrum. Meas.}\ }\textbf
  {\bibinfo {volume} {52}},\ \bibinfo {pages} {524} (\bibinfo {year}
  {2003})}\BibitemShut {NoStop}%
\bibitem [{\citenamefont {Delahaye}\ and\ \citenamefont
  {Jeckelmann}(2003)}]{Delahaye2003}%
  \BibitemOpen
  \bibfield  {author} {\bibinfo {author} {\bibfnamefont {F.}~\bibnamefont
  {Delahaye}}\ and\ \bibinfo {author} {\bibfnamefont {B.}~\bibnamefont
  {Jeckelmann}},\ }\bibfield  {title} {\enquote {\bibinfo {title} {Revised
  technical guidelines for reliable dc measurements of the quantized hall
  resistance},}\ }\href@noop {} {\bibfield  {journal} {\bibinfo  {journal}
  {Metrologia}\ }\textbf {\bibinfo {volume} {40}},\ \bibinfo {pages} {217}
  (\bibinfo {year} {2003})}\BibitemShut {NoStop}%
\bibitem [{\citenamefont {Delahaye}(1993)}]{Delahaye1993}%
  \BibitemOpen
  \bibfield  {author} {\bibinfo {author} {\bibfnamefont {F.}~\bibnamefont
  {Delahaye}},\ }\bibfield  {title} {\enquote {\bibinfo {title} {Series and
  parallel connection of multiterminal quantum hall-effect devices},}\
  }\href@noop {} {\bibfield  {journal} {\bibinfo  {journal} {J. Appl. Phys.}\
  }\textbf {\bibinfo {volume} {73}},\ \bibinfo {pages} {7914} (\bibinfo {year}
  {1993})}\BibitemShut {NoStop}%
\bibitem [{\citenamefont {Poirier}\ \emph {et~al.}(2014)\citenamefont
  {Poirier}, \citenamefont {Lafont}, \citenamefont {Djordjevic}, \citenamefont
  {Schopfer},\ and\ \citenamefont {Devoille}}]{Poirier2014}%
  \BibitemOpen
  \bibfield  {author} {\bibinfo {author} {\bibfnamefont {W.}~\bibnamefont
  {Poirier}}, \bibinfo {author} {\bibfnamefont {F.}~\bibnamefont {Lafont}},
  \bibinfo {author} {\bibfnamefont {S.}~\bibnamefont {Djordjevic}}, \bibinfo
  {author} {\bibfnamefont {F.}~\bibnamefont {Schopfer}}, \ and\ \bibinfo
  {author} {\bibfnamefont {L.}~\bibnamefont {Devoille}},\ }\bibfield  {title}
  {\enquote {\bibinfo {title} {A programmable quantum current standard from the
  josephson and the quantum hall effects},}\ }\href@noop {} {\bibfield
  {journal} {\bibinfo  {journal} {J. Appl. Phys.}\ }\textbf {\bibinfo {volume}
  {115}},\ \bibinfo {pages} {044509} (\bibinfo {year} {2014})}\BibitemShut
  {NoStop}%
\bibitem [{\citenamefont {Harvey}(1972)}]{Harvey1972}%
  \BibitemOpen
  \bibfield  {author} {\bibinfo {author} {\bibfnamefont {I.~K.}\ \bibnamefont
  {Harvey}},\ }\bibfield  {title} {\enquote {\bibinfo {title} {A precise low
  temperature dc ratio transformer},}\ }\href@noop {} {\bibfield  {journal}
  {\bibinfo  {journal} {Rev. Sci. Instrum.}\ }\textbf {\bibinfo {volume}
  {43}},\ \bibinfo {pages} {1626} (\bibinfo {year} {1972})}\BibitemShut
  {NoStop}%
\bibitem [{\citenamefont {Witt}(2005)}]{Witt2005}%
  \BibitemOpen
  \bibfield  {author} {\bibinfo {author} {\bibfnamefont {T.}~\bibnamefont
  {Witt}},\ }\bibfield  {title} {\enquote {\bibinfo {title} {Allan variances
  and spectral densities for dc voltage measurements with polarity
  reversals},}\ }\href@noop {} {\bibfield  {journal} {\bibinfo  {journal} {IEEE
  Trans. Instrum. Meas.}\ }\textbf {\bibinfo {volume} {54}},\ \bibinfo {pages}
  {550} (\bibinfo {year} {2005})}\BibitemShut {NoStop}%
\bibitem [{\citenamefont {Kibble}(2010)}]{Kibble2010}%
  \BibitemOpen
  \bibfield  {author} {\bibinfo {author} {\bibfnamefont {B.~P.}\ \bibnamefont
  {Kibble}},\ }\bibfield  {title} {\enquote {\bibinfo {title} {In metrology,
  simpler is better},}\ }\href@noop {} {\bibfield  {journal} {\bibinfo
  {journal} {IEEE Intrum. Meas. Mag.}\ }\textbf {\bibinfo {volume} {13}},\
  \bibinfo {pages} {43} (\bibinfo {year} {2010})}\BibitemShut {NoStop}%
\bibitem [{\citenamefont {Benz}\ \emph {et~al.}(2015)\citenamefont {Benz},
  \citenamefont {Waltman}, \citenamefont {Fox}, \citenamefont {Dresselhaus},
  \citenamefont {Rüfenacht}, \citenamefont {Underwood}, \citenamefont {Howe},
  \citenamefont {Schwall},\ and\ \citenamefont {Burroughs}}]{Benz2015}%
  \BibitemOpen
  \bibfield  {author} {\bibinfo {author} {\bibfnamefont {S.~P.}\ \bibnamefont
  {Benz}}, \bibinfo {author} {\bibfnamefont {S.~B.}\ \bibnamefont {Waltman}},
  \bibinfo {author} {\bibfnamefont {A.~E.}\ \bibnamefont {Fox}}, \bibinfo
  {author} {\bibfnamefont {P.~D.}\ \bibnamefont {Dresselhaus}}, \bibinfo
  {author} {\bibfnamefont {A.}~\bibnamefont {Rüfenacht}}, \bibinfo {author}
  {\bibfnamefont {J.~M.}\ \bibnamefont {Underwood}}, \bibinfo {author}
  {\bibfnamefont {L.~A.}\ \bibnamefont {Howe}}, \bibinfo {author}
  {\bibfnamefont {R.~E.}\ \bibnamefont {Schwall}}, \ and\ \bibinfo {author}
  {\bibfnamefont {C.~J.}\ \bibnamefont {Burroughs}},\ }\bibfield  {title}
  {\enquote {\bibinfo {title} {One-volt josephson arbitrary waveform
  synthesizer},}\ }\href@noop {} {\bibfield  {journal} {\bibinfo  {journal}
  {IEEE Trans. Appl. Supercond.}\ }\textbf {\bibinfo {volume} {25}},\ \bibinfo
  {pages} {1300108} (\bibinfo {year} {2015})}\BibitemShut {NoStop}%
\bibitem [{\citenamefont {Kieler}\ \emph {et~al.}(2015)\citenamefont {Kieler},
  \citenamefont {Behr}, \citenamefont {Wendisch}, \citenamefont {Bauer},
  \citenamefont {Palafox},\ and\ \citenamefont {Kohlmann}}]{Kieler2015}%
  \BibitemOpen
  \bibfield  {author} {\bibinfo {author} {\bibfnamefont {O.~F.}\ \bibnamefont
  {Kieler}}, \bibinfo {author} {\bibfnamefont {R.}~\bibnamefont {Behr}},
  \bibinfo {author} {\bibfnamefont {R.}~\bibnamefont {Wendisch}}, \bibinfo
  {author} {\bibfnamefont {S.}~\bibnamefont {Bauer}}, \bibinfo {author}
  {\bibfnamefont {L.}~\bibnamefont {Palafox}}, \ and\ \bibinfo {author}
  {\bibfnamefont {J.}~\bibnamefont {Kohlmann}},\ }\bibfield  {title} {\enquote
  {\bibinfo {title} {Towards a 1 v josephson arbitrary waveform synthesizer},}\
  }\href@noop {} {\bibfield  {journal} {\bibinfo  {journal} {IEEE Trans. Appl.
  Supercond.}\ }\textbf {\bibinfo {volume} {25}},\ \bibinfo {pages} {1400305}
  (\bibinfo {year} {2015})}\BibitemShut {NoStop}%
\bibitem [{\citenamefont {Ahlers}\ \emph {et~al.}(2009)\citenamefont {Ahlers},
  \citenamefont {Jeannneret}, \citenamefont {Overney}, \citenamefont {Schurr},\
  and\ \citenamefont {Wood}}]{Ahlers2009}%
  \BibitemOpen
  \bibfield  {author} {\bibinfo {author} {\bibfnamefont {F.~J.}\ \bibnamefont
  {Ahlers}}, \bibinfo {author} {\bibfnamefont {B.}~\bibnamefont {Jeannneret}},
  \bibinfo {author} {\bibfnamefont {F.}~\bibnamefont {Overney}}, \bibinfo
  {author} {\bibfnamefont {J.}~\bibnamefont {Schurr}}, \ and\ \bibinfo {author}
  {\bibfnamefont {B.~M.}\ \bibnamefont {Wood}},\ }\bibfield  {title} {\enquote
  {\bibinfo {title} {Compendium for precise ac measurements of the quantized
  hall resistance},}\ }\href@noop {} {\bibfield  {journal} {\bibinfo  {journal}
  {Metrologia}\ }\textbf {\bibinfo {volume} {46}},\ \bibinfo {pages} {R1}
  (\bibinfo {year} {2009})}\BibitemShut {NoStop}%
\bibitem [{\citenamefont {Kibble}\ and\ \citenamefont
  {Rayner}(1984)}]{Kibble1984}%
  \BibitemOpen
  \bibfield  {author} {\bibinfo {author} {\bibfnamefont {B.~P.}\ \bibnamefont
  {Kibble}}\ and\ \bibinfo {author} {\bibfnamefont {G.~H.}\ \bibnamefont
  {Rayner}},\ }\href@noop {} {\emph {\bibinfo {title} {Coaxial AC Bridges}}}\
  (\bibinfo  {publisher} {Adam Hilger Ltd, Bristol},\ \bibinfo {year}
  {1984})\BibitemShut {NoStop}%
\bibitem [{\citenamefont {Scherer}\ and\ \citenamefont
  {Camarota}(2012)}]{Scherer2012}%
  \BibitemOpen
  \bibfield  {author} {\bibinfo {author} {\bibfnamefont {H.}~\bibnamefont
  {Scherer}}\ and\ \bibinfo {author} {\bibfnamefont {B.}~\bibnamefont
  {Camarota}},\ }\bibfield  {title} {\enquote {\bibinfo {title} {Quantum
  metrology triangle experiments: a status review},}\ }\href@noop {} {\bibfield
   {journal} {\bibinfo  {journal} {Meas. Sci. Technolo.}\ }\textbf {\bibinfo
  {volume} {23}},\ \bibinfo {pages} {124010} (\bibinfo {year}
  {2012})}\BibitemShut {NoStop}%
\bibitem [{\citenamefont {Jehl}\ \emph {et~al.}(2013)\citenamefont {Jehl},
  \citenamefont {Voisin}, \citenamefont {Charron}, \citenamefont {Clapera},
  \citenamefont {Ray}, \citenamefont {Roche}, \citenamefont {Sanquer},
  \citenamefont {Djordjevic}, \citenamefont {Devoille}, \citenamefont
  {Wacquez},\ and\ \citenamefont {Vinet}}]{Jehl2013}%
  \BibitemOpen
  \bibfield  {author} {\bibinfo {author} {\bibfnamefont {X.}~\bibnamefont
  {Jehl}}, \bibinfo {author} {\bibfnamefont {B.}~\bibnamefont {Voisin}},
  \bibinfo {author} {\bibfnamefont {T.}~\bibnamefont {Charron}}, \bibinfo
  {author} {\bibfnamefont {P.}~\bibnamefont {Clapera}}, \bibinfo {author}
  {\bibfnamefont {S.}~\bibnamefont {Ray}}, \bibinfo {author} {\bibfnamefont
  {B.}~\bibnamefont {Roche}}, \bibinfo {author} {\bibfnamefont
  {M.}~\bibnamefont {Sanquer}}, \bibinfo {author} {\bibfnamefont
  {S.}~\bibnamefont {Djordjevic}}, \bibinfo {author} {\bibfnamefont
  {L.}~\bibnamefont {Devoille}}, \bibinfo {author} {\bibfnamefont
  {R.}~\bibnamefont {Wacquez}}, \ and\ \bibinfo {author} {\bibfnamefont
  {M.}~\bibnamefont {Vinet}},\ }\bibfield  {title} {\enquote {\bibinfo {title}
  {Hybrid metal-semiconductor electron pump for quantum metrology},}\
  }\href@noop {} {\bibfield  {journal} {\bibinfo  {journal} {Phys. Rev. X}\
  }\textbf {\bibinfo {volume} {3}} (\bibinfo {year} {2013})}\BibitemShut
  {NoStop}%
\bibitem [{\citenamefont {Devoille}\ \emph {et~al.}(2012)\citenamefont
  {Devoille}, \citenamefont {Feltin}, \citenamefont {Steck}, \citenamefont
  {Chenaud}, \citenamefont {Sassine}, \citenamefont {Djordevic}, \citenamefont
  {S\'eron},\ and\ \citenamefont {Piquemal}}]{Devoille2012}%
  \BibitemOpen
  \bibfield  {author} {\bibinfo {author} {\bibfnamefont {L.}~\bibnamefont
  {Devoille}}, \bibinfo {author} {\bibfnamefont {N.}~\bibnamefont {Feltin}},
  \bibinfo {author} {\bibfnamefont {B.}~\bibnamefont {Steck}}, \bibinfo
  {author} {\bibfnamefont {B.}~\bibnamefont {Chenaud}}, \bibinfo {author}
  {\bibfnamefont {S.}~\bibnamefont {Sassine}}, \bibinfo {author} {\bibfnamefont
  {S.}~\bibnamefont {Djordevic}}, \bibinfo {author} {\bibfnamefont
  {O.}~\bibnamefont {S\'eron}}, \ and\ \bibinfo {author} {\bibfnamefont
  {F.}~\bibnamefont {Piquemal}},\ }\bibfield  {title} {\enquote {\bibinfo
  {title} {Quantum metrological triangle experiment at lne: measurements on a
  three-junction r-pump using a 20000:1 winding ratio cryogenic current
  comparator},}\ }\href@noop {} {\bibfield  {journal} {\bibinfo  {journal}
  {Meas. Sci. Technolo.}\ }\textbf {\bibinfo {volume} {23}},\ \bibinfo {pages}
  {124011} (\bibinfo {year} {2012})}\BibitemShut {NoStop}%
\bibitem [{\citenamefont {BIPM}(2008)}]{GUM}%
  \BibitemOpen
  \bibfield  {author} {\bibinfo {author} {\bibnamefont {BIPM}},\ }\href@noop {}
  {\emph {\bibinfo {title} {Evaluation of measurement data — Guide to the
  expression of uncertainty in measurement}}}\ (\bibinfo  {publisher} {JCGM},\
  \bibinfo {year} {2008})\BibitemShut {NoStop}%
\bibitem [{\citenamefont {Piquemal}\ \emph {et~al.}(1993)\citenamefont
  {Piquemal}, \citenamefont {Genev\`es}, \citenamefont {Delahaye},
  \citenamefont {Andr\'e}, \citenamefont {Patillon},\ and\ \citenamefont
  {Frijlink}}]{Piquemal1993}%
  \BibitemOpen
  \bibfield  {author} {\bibinfo {author} {\bibfnamefont {F.}~\bibnamefont
  {Piquemal}}, \bibinfo {author} {\bibfnamefont {G.}~\bibnamefont {Genev\`es}},
  \bibinfo {author} {\bibfnamefont {F.}~\bibnamefont {Delahaye}}, \bibinfo
  {author} {\bibfnamefont {J.~P.}\ \bibnamefont {Andr\'e}}, \bibinfo {author}
  {\bibfnamefont {J.~N.}\ \bibnamefont {Patillon}}, \ and\ \bibinfo {author}
  {\bibfnamefont {P.}~\bibnamefont {Frijlink}},\ }\bibfield  {title} {\enquote
  {\bibinfo {title} {Report on a joint bipm-euromet project for the fabrication
  of qhe samples by the lep},}\ }\href@noop {} {\bibfield  {journal} {\bibinfo
  {journal} {IEEE Trans. Instrum. Meas.}\ }\textbf {\bibinfo {volume} {42}},\
  \bibinfo {pages} {264} (\bibinfo {year} {1993})}\BibitemShut {NoStop}%
\bibitem [{\citenamefont {Lafont}\ \emph {et~al.}(2015)\citenamefont {Lafont},
  \citenamefont {Ribeiro-Palau}, \citenamefont {Kazazis}, \citenamefont
  {Michon}, \citenamefont {Couturaud}, \citenamefont {Consejo}, \citenamefont
  {Chassagne}, \citenamefont {Zielinski}, \citenamefont {Portail},
  \citenamefont {Jouault}, \citenamefont {Schopfer},\ and\ \citenamefont
  {Poirier}}]{Lafont2015}%
  \BibitemOpen
  \bibfield  {author} {\bibinfo {author} {\bibfnamefont {F.}~\bibnamefont
  {Lafont}}, \bibinfo {author} {\bibfnamefont {R.}~\bibnamefont
  {Ribeiro-Palau}}, \bibinfo {author} {\bibfnamefont {D.}~\bibnamefont
  {Kazazis}}, \bibinfo {author} {\bibfnamefont {A.}~\bibnamefont {Michon}},
  \bibinfo {author} {\bibfnamefont {O.}~\bibnamefont {Couturaud}}, \bibinfo
  {author} {\bibfnamefont {C.}~\bibnamefont {Consejo}}, \bibinfo {author}
  {\bibfnamefont {T.}~\bibnamefont {Chassagne}}, \bibinfo {author}
  {\bibfnamefont {M.}~\bibnamefont {Zielinski}}, \bibinfo {author}
  {\bibfnamefont {M.}~\bibnamefont {Portail}}, \bibinfo {author} {\bibfnamefont
  {B.}~\bibnamefont {Jouault}}, \bibinfo {author} {\bibfnamefont
  {F.}~\bibnamefont {Schopfer}}, \ and\ \bibinfo {author} {\bibfnamefont
  {W.}~\bibnamefont {Poirier}},\ }\bibfield  {title} {\enquote {\bibinfo
  {title} {Quantum hall resistance standards from graphene grown by chemical
  vapour deposition on silicon carbide},}\ }\href@noop {} {\bibfield  {journal}
  {\bibinfo  {journal} {Nature Communications}\ }\textbf {\bibinfo {volume}
  {6}},\ \bibinfo {pages} {6806} (\bibinfo {year} {2015})}\BibitemShut
  {NoStop}%
\bibitem [{\citenamefont {Ricketts}\ and\ \citenamefont
  {Kemeny}(1988)}]{Ricketts1988}%
  \BibitemOpen
  \bibfield  {author} {\bibinfo {author} {\bibfnamefont {B.~W.}\ \bibnamefont
  {Ricketts}}\ and\ \bibinfo {author} {\bibfnamefont {P.~C.}\ \bibnamefont
  {Kemeny}},\ }\bibfield  {title} {\enquote {\bibinfo {title} {Quantum hall
  effect devices as circuit elements},}\ }\href@noop {} {\bibfield  {journal}
  {\bibinfo  {journal} {J. Phys. D}\ }\textbf {\bibinfo {volume} {21}},\
  \bibinfo {pages} {483} (\bibinfo {year} {1988})}\BibitemShut {NoStop}%
\end{thebibliography}
\providecommand{\noopsort}[1]{}\providecommand{\singleletter}[1]{#1}%

\end{document}